\documentclass[11pt]{article}
  \pdfoutput=1
\usepackage{jheppub}

\usepackage{amsmath}
\usepackage{verbatim}   
\usepackage{subfigure}  
\usepackage{amsfonts}
\usepackage{wasysym}
\usepackage{amssymb}
\usepackage{mathrsfs}
\usepackage{graphicx}
 \usepackage{slashed}
 \usepackage{epsfig}
 \usepackage{url}
\usepackage{color}

\usepackage{epigraph}

 \graphicspath{{figures/}}

\newcommand{\newc}{\newcommand}
\newc{\gsim}{\lower.7ex\hbox{$\;\stackrel{\textstyle>}{\sim}\;$}}
\newc{\lsim}{\lower.7ex\hbox{$\;\stackrel{\textstyle<}{\sim}\;$}}
\newc{\gev}{\,{\rm GeV}}
\newc{\mev}{\,{\rm MeV}}
\newc{\ev}{\,{\rm eV}}
\newc{\kev}{\,{\rm keV}}
\newc{\tev}{\,{\rm TeV }}

\def\Im{\mathop{\rm Im}}
\def\Re{\mathop{\rm Re}}

\newc{\mz}{M_Z}
\newc{\mpl}{M_*}
\newc{\mw}{m_{\rm weak}}
\newc{\nr}[1]{N^c_R{}_{#1}}
\usepackage{amsmath}

\def\beq{\begin{equation}}
\def\eeq{\end{equation}}
\def\bea{\begin{eqnarray}}
\def\eea{\end{eqnarray}}
\def\bitem{\begin{itemize}}
\def\eitem{\end{itemize}}
\newcommand{\bec}{\begin{center}}
\newcommand{\eec}{\end{center}}
\newcommand{\half}{\frac{1}{2}}

     \newcommand{\fb}{{\mathrm {fb}}}

\def\bar#1{\overline{#1}}
\def\vev#1{\left\langle #1 \right\rangle}
\def\bra#1{\left\langle #1\right|}
\def\ket#1{\left| #1\right\rangle}

\def\inv{^{\raise.15ex\hbox{${\scriptscriptstyle -}$}\kern-.05em 1}}
\def\lbar{{\lower.35ex\hbox{$\mathchar'26$}\mkern-10mu\lambda}} 

\def\to{\rightarrow}

\let\<=\langle
\let\>=\rangle

\let\+=\uparrow

\def\bmu{B_{\mu}}

\def\slashchar#1{\ensuremath{                               %
   \setbox0=\hbox{${}#1{}$}       
   \dimen0=\wd0                                 
   \setbox1=\hbox{/} \dimen1=\wd1               
   \ifdim\dimen0>\dimen1                        
      \rlap{\hbox to \dimen0{\hfil/\hfil}}      
      {}#1{}                                    
   \else                                        
      \rlap{\hbox to \dimen1{\hfil${}#1{}$\hfil}}   
   \fi}} 

\begin{document}
\hfill \vspace{-5mm} SITP-13/16

\title{The Last Vestiges of Naturalness}

\author[a]{Asimina Arvanitaki,}
\emailAdd{aarvan@stanford.edu}

\author[a]{Masha Baryakhtar,}
\emailAdd{mbaryakh@stanford.edu}

\author[a]{Xinlu Huang,}
\emailAdd{xinluh@stanford.edu}

\author[a]{Ken~Van~Tilburg,}
\emailAdd{kenvt@stanford.edu}

\author[a,b]{and Giovanni~Villadoro}
\emailAdd{giovanni.villadoro@ictp.it}
\affiliation[a]{Stanford Institute for Theoretical Physics, Department of Physics,\\
 Stanford University, Stanford, CA 94305, USA}
\affiliation[b]{Abdus Salam International Centre for Theoretical Physics\\
Strada Costiera 11, 34151, Trieste, Italy}
%

\date{\today}

\abstract{Direct LHC bounds on colored SUSY particles now corner
  naturalness more than the measured value of the Higgs mass
  does. Bounds on the gluino are of particular importance, since it
  radiatively ``sucks" up the stop and Higgs-up soft masses. As a
  result, even models that easily accommodate a 125 GeV Higgs are
  almost as tuned as the simplest version of SUSY, the MSSM: at best
  at the percent level. In this paper, we further examine how current
  LHC results constrain naturalness in three classes of models that
  may relax LHC bounds on sparticles: split families, baryonic
  RPV, and Dirac gauginos. In models of split families and bRPV, the
  bounds on the gluino are only slightly reduced, resulting in a few
  percent tuning. In particular, having a natural spectrum in bRPV
  models typically implies that tops, $W$s, and $Z$s are easily produced
  in the cascade decays of squarks and gluinos. The resulting leptons
  and missing energy push the gluino mass limit above 1 TeV. Even when
  the gluino has a Dirac mass and does not contribute to the stop mass
  at one loop, tuning reappears in calculable models because there is
  no symmetry imposing the supersoft limit. We conclude that, even if
  sparticles are found at LHC-14, naturalness will not emerge
  triumphant. }

\maketitle

\section{Introduction}
\label{sec:intro}

The LHC has brought particle physics face to face with data. 
It discovered the last missing particle of the Standard Model, the
Higgs boson, and it is placing increasingly stringent bounds on the
new physics proposed to protect its mass. The principle of
Naturalness, which has been guiding beyond the Standard Model (BSM)
physics for several decades, is now in question, and the possibility
that the electroweak (EW) scale is tuned by environmental selection in
a multiverse looks increasingly alluring.

Already after the LEP experiments, the absence of any BSM particle at or below the electroweak scale suggested the presence of a small gap between the $Z$ mass and the scale of new physics.
In fact, the principle of Naturalness has already failed once:
the smallness of the cosmological constant (CC) produces a hierarchy at least 50 orders of magnitude more severe, already a strong indication for environmental selection in our universe. 

However, unlike for the CC problem, many dynamical solutions for stabilizing the EW scale exist. Among these, supersymmetry (SUSY) plays a special role because it quantitatively predicts the unification of the gauge couplings \cite{Dimopoulos:1981yj,Dimopoulos:1981zb}. The latter gives an independent, although weaker, motivation to the presence of new particles around the TeV scale \cite{ArkaniHamed:2004fb}.

It is well-known that in the minimal implementation of SUSY---the
MSSM---raising the Higgs mass to its experimental value of $125~\gev$
requires stop masses above a TeV, already implying a tuning at the
percent level or worse (depending on $A$-terms). Such large radiative
corrections can be avoided if there are extra contributions to the Higgs quartic self-coupling, as in the NMSSM, in which a singlet provides this contribution. However, direct bounds on colored superpartners from the LHC are already so strong that even after raising the value of the Higgs mass in a natural way, the fine tuning does not improve. Thus, the only hope for finding a natural implementations of SUSY is to look for models that can relax the most stringent bounds set by the LHC searches. Recently, three classes of models have attracted special attention: split families (also dubbed ``Natural'' SUSY), baryonic R-parity violating SUSY (bRPV), and Dirac gauginos.

The structure of the paper is as follows: in the remainder of this section, we discuss the sources of tuning in SUSY theories after the first run of the LHC. In particular, the MSSM and the NMSSM constitute a benchmark for the other models. In sections~\ref{sec:splitfam},~\ref{sec:rpv} and \ref{sec:dirac}, we analyze split-family, bRPV, and Dirac gaugino models, respectively.  We discuss experimental constraints and compare the fine tuning in explicit models wherever possible. In section~\ref{sec:conclusion}, we summarize our conclusions about the various models and assess the status of naturalness in SUSY.

\subsection{Quantifying Naturalness}
\label{sec:QN}
The level of tuning in a theory can be measured by the sensitivity of the low-energy observables $\mathcal{O}$ to the fundamental UV parameters $a_i$; it is commonly quantified by the formula \cite{Barbieri:1987fn}:
\begin{eqnarray} \label{eq:deftune1}
\text{FT}_{{\cal O}}=\left[ \sum_i {\left(\frac {\partial \log {\cal O}}{\partial \log a_i}\right)^2} \right]^{-1/2}\,.
\end{eqnarray}
The total amount of fine tuning of a theory is given by the product of the fine tunings of each
independent low-energy observable,
\begin{equation} \label{eq:deftune2}
\text{FT}=\prod_i \text{FT}_{{\cal O}_i}\,.
\end{equation}
For theories that try to address the hierarchy problem, the dominant source of fine tuning is usually the one associated to the EW scale. There are cases in which the tuning resides in other observables; we take these into account with eq.~(\ref{eq:deftune2}).\footnote{When two observables are not independent one should take into account the degree of correlation to avoid overestimating the tuning (details in appendix~\ref{app:tuning}).} 

In the decoupling limit for the lightest Higgs state $h$, the tree-level effective potential is
\begin{equation}
V(h)=\frac{1}{2} m_0^2 h^2+\frac14 \lambda h^4\,.
\end{equation}
At the minimum $\langle h \rangle=v$, the physical Higgs mass $m_h$, the vev and the Lagrangian mass term $m_0$
are related by
\begin{equation}
m_h^2=2\lambda v^2 =-2m_0^2\,.
\end{equation} 
It is clear that in this case the tuning for $v^2$ and for $m_h^2$ are
equal, and can be computed independently of the quartic coupling and its origin.

In the MSSM at large $\tan \beta$,
\begin{equation}
m_h^2 = - 2m_0^2=-2\left(m_{H_u}^2+|\mu|^2\right),\label{eq:mhiggslargetb}
\end{equation}
which implies that the higgsino cannot be much heavier than the lightest Higgs without tree-level tuning. 
From the one loop RGE, 
\begin{equation}
\partial_{t} m_{H_u}^2=\frac{6|y_t|^2}{(4\pi)^2}\left( m_{\tilde t_L}^2+m_{\tilde t_R}^2+|A_t|^2\right)+\dots,\label{eq:stoprunning}
\end{equation}
we see that $m_{H_u}^2$ receives large corrections proportional to the
stop masses and the $A$-term, which thus need to be small as well. Finally, the stop mass itself is attracted to the gluino mass in the IR,
\begin{equation}
\partial_t m_{\tilde t}^2=-\frac{8\alpha_s}{3\pi} M_{3}^2+\dots,\label{eq:gluinorunning}
\end{equation}
which indirectly attracts $m_{H_u}^2$. Because of the large coefficients in
eqs.~(\ref{eq:stoprunning}) and (\ref{eq:gluinorunning}) and the strong
experimental bounds, the gluino contribution is typically the dominant
source of tuning.  Note that to capture non-linear effects in the
RGEs, we implement a full-scale RG analysis with tuning against UV
parameters, instead of just using the leading-log estimates.
The magnitude of the RG contributions to the Higgs mass from the stop
and the gluino increases with the amount of running; therefore models with a low mediation scale are typically less tuned. 

While eq.~(\ref{eq:mhiggslargetb}) is exactly true in the
large-$\tan\beta$ limit of the MSSM, we use the same formula as a
reliable estimate at lower values of $\tan\beta$ and in more general
Higgs sectors. Relaxing the large-$\tan\beta$ assumption does not
alleviate the tuning because in the decoupling limit the physical
Higgs always couples to the stop with the physical top Yukawa coupling ($m_t/v$), independently of $\tan\beta$.\footnote{At $\tan \beta \approx 1$, the tuning is moved from the minimization condition in eq.~(\ref{eq:mhiggslargetb}) to the requirement $2\bmu~\simeq~\left(m_{H_u}^2 + m_{H_d}^2 + 2|\mu|^2\right)$, again imposing an additive relation on mass parameters.}
In extended Higgs sectors, the tuning can be reduced by changing the
relation between the measured Higgs mass and the vev of
eq.~(\ref{eq:mhiggslargetb}). However, since the $125\gev$ resonance
couples to vector bosons with very SM-like couplings (up to $\sim
25\%$ \cite{Chatrchyan:2013lba,Aad:2013wqa}), it should be the one mostly responsible for EW symmetry breaking and thus for most of the top mass. For example, in the MSSM the mixing between the two Higgs doublets is already constrained to be below 10\% \cite{craigsetal}. This implies a significant coupling to the stop, which causes the tuning.
On the other hand, in the NMSSM it is possible to adjust the parameters to simulate SM couplings for a 125~GeV resonance that is not the state mostly responsible for EWSB. In this case, the tuning of the EW vev is shifted to tuning the values of the Higgs couplings to be SM-like. Large mixings generically imply big deviations of the Higgs couplings from their SM values---a natural Higgs is not the SM Higgs \cite{Arvanitaki:2011ck}.

\begin{figure}[t]
\begin{center}
   {\includegraphics[trim = 0mm 0mm 0mm 0mm, clip, width
      = 0.65\textwidth]{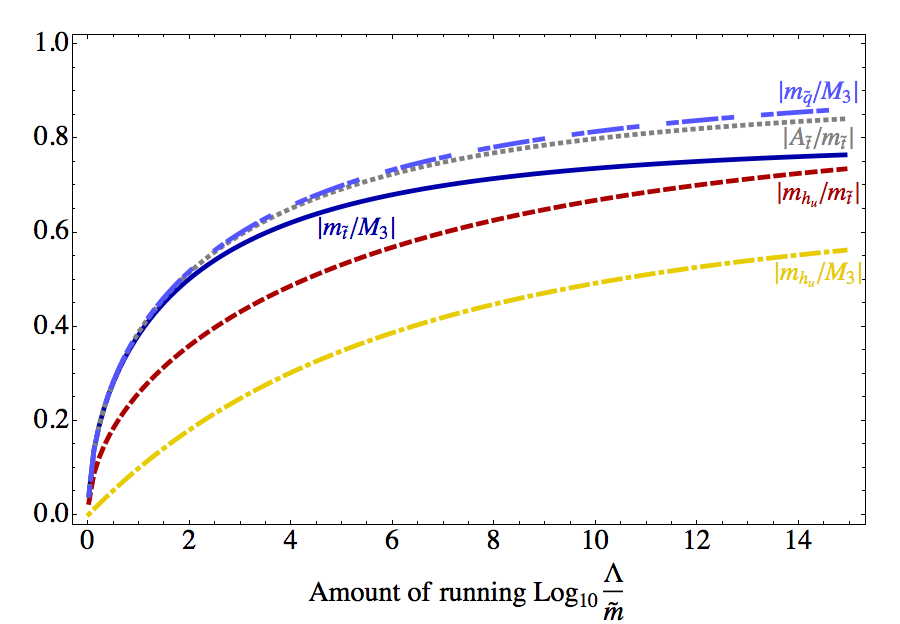}}
\vspace{-2mm}
\caption{\label{fig:gluinosucks}  The \emph{gluino sucks} effect: Even starting with vanishing boundary conditions for all scalar soft terms at the scale $\Lambda$, they are quickly generated in the IR by the gluino mass contributions. Already after one decade of running the average stop mass $m_{\tilde t}=\sqrt{(m_{{\tilde t}_L}^2+m_{{\tilde t}_R}^2)/2}$ is almost a factor 2 below the gluino and after three decades also $|m_{H_u}|$ is within a factor of 4 from the gluino. Few decades of running are enough for the soft masses to saturate their IR fixed values. 
}
   \end{center}
\end{figure}

One may think that a light stop, regardless of the gluino mass, improves the $m_h^2$ tuning by reducing the one-loop contribution in eq.~(\ref{eq:stoprunning}). On the contrary, this scenario is even more tuned because there are \emph{two} unnaturally light scalars.  The large coefficient in front of $M_3^2$ in eq.~(\ref{eq:gluinorunning}) imposes a IR fixed relation  between $m_{\tilde t}$ and $M_3$, and the least tuned spectra are those where $m_{\tilde t}\sim M_3$. As shown in fig.~\ref{fig:gluinosucks}, both $m_{\tilde t}$ and $m_{H_u}$ are quickly attracted by the gluino even if they vanish at the messenger scale. Deviating from this IR prediction requires more tuning. Models with a low  mediation scale are thus preferred.

In order to significantly relax the bounds on Naturalness, one has to
alleviate the strong ``gluino sucks" effect. We can envision two
possible ways to accomplish this: hiding the gluino (by relaxing the
collider bounds) or canceling the gluino effects with other big
contributions. The first class of solutions includes split family
models (which decrease the production cross section) and bRPV models
(which hide the collider signatures). The second class includes Dirac
gaugino models (where contributions from extra partner fields cancel
the gluino ones). All these models, which are discussed in detail
later, are based on symmetries or a dynamical structure. Their
beneficial properties are robust against the UV details of the theory,
thus giving them an opportunity to alleviate the tuning.

The models that may improve naturalness do not include those that
suppress the RG effects of the gluino without a symmetry or a robust
dynamical mechanism. These include stealth/compressed spectra models
where the gluino bounds are relaxed by means of a ``cleverly
arranged'' low-energy spectrum, and focus-point-like models where the gluino RG contributions are canceled 
against other enhanced contributions (sourced by the other gauginos, squarks, etc.) 
via accurately chosen boundary conditions at high energies. In all these
cases there is no symmetry imposing the UV relations, and the tuning
is shifted to the carefully chosen boundary conditions. 
The presence of these accidental cancellations in the RGE and the reason why
they are not a solution of the tuning problem was already explained in the seminal
paper by Barbieri and Giudice \cite{Barbieri:1987fn} well before the more
recent revival after the LEP, Tevatron and LHC negative results.
Also note that such tuning cannot be avoided by replacing the tuned UV parameters with
discrete ones.  Even if the standard formula in
eq.~(\ref{eq:deftune1}) does not reflect this kind of tuning, the
model building assumptions generate a mini-landscape, where only a
small fraction of models satisfy the required UV conditions. In fact,
these kinds of models are implementing a solution to the hierarchy
problem similar in spirit to the string landscape: after moduli
stabilization, string theory vacua are determined by choices of
discrete parameters (topology, flux units, number of branes, etc.) and
each ``model'' has no free continuous parameter, thus no tuning
according to eq.~(\ref{eq:deftune1}). This example highlights the
danger of blindly applying tuning formulae such as
eq.~(\ref{eq:deftune1}) while practicing extreme model building.

In the remainder of this paper, we study low-scale gauge mediated (GMSB) models, which minimize the effects of RG flows, automatically address the flavor problem, preserve unification, and provide a calculable framework for computing the tuning. We minimize the boundary contribution to the scalar masses by taking a large number of messengers and allow extra contributions to the stop and Higgs soft masses at the UV as suggested by models solving the $\mu$-$\bmu$ problem~\cite{Giudice:1998bp,Giudice:2007ca,DeSimone:2011va}. 
In computing the tuning according to eqs.~(\ref{eq:deftune1}) and (\ref{eq:deftune2}), we use the observables $\mathcal{O}_i = \{ m_h^2 , m_{\tilde{t}}^2\}$ evaluated at the scale of the stop masses and with $m_h^2$ as in eq.~(\ref{eq:mhiggslargetb}). We also use different sets of the UV parameters $a_i$, which depend on the explicit model and will be specified in the appropriate sections. This procedure gives us a reliable framework to compare the tuning across models. We expect our explicit models to exhibit nearly optimal tuning.

\subsection{Setting the Standard for minimal fine-tuning in the MSSM}\label{sec:settingstandard}
\begin{figure}[t]
\begin{center}
\vspace{-4mm}
{ \large \emph{(a)} \bf MSSM} \\ 
   \includegraphics[trim = 0mm 0mm 0mm 0mm, clip, width
      = 0.70\textwidth]{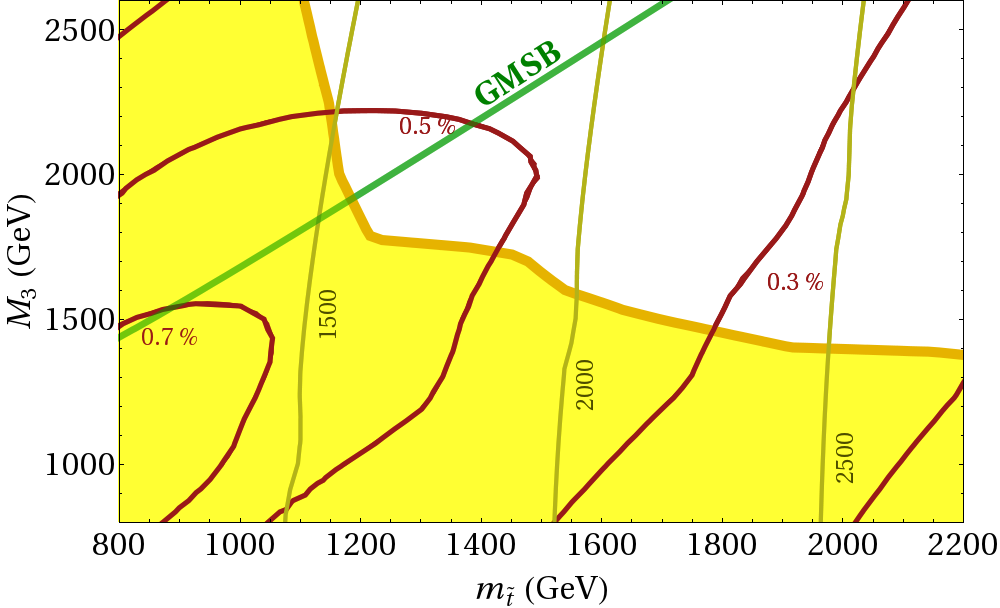}
\\[.1cm]
{ \large \emph{(b)} \bf NMSSM} \\ 
  \includegraphics[trim = 0mm 0mm 0mm 0mm, clip, width
      = 0.70\textwidth]{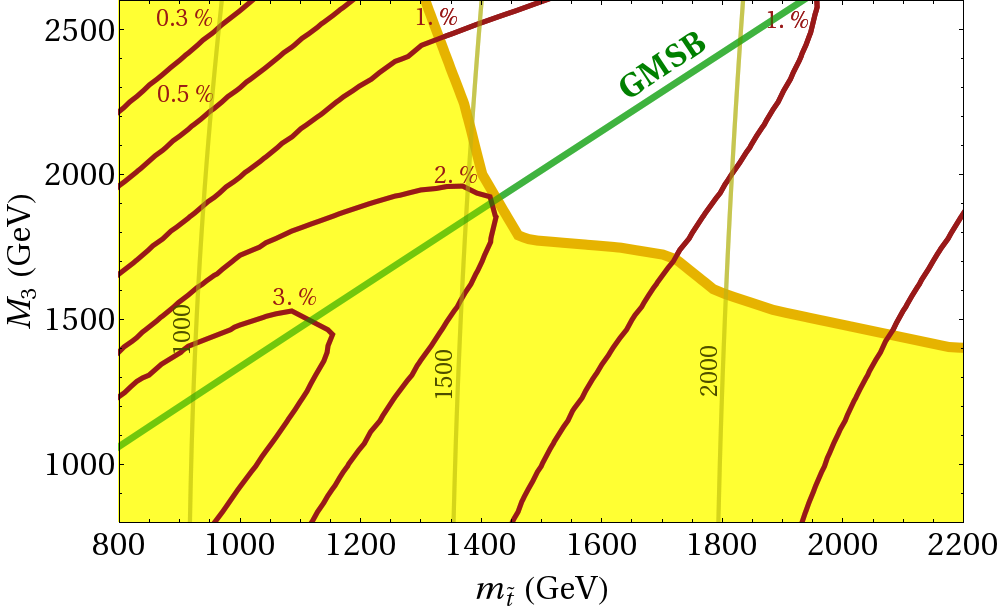}
\vspace{-3mm}
\caption{ \label{fig:pushup}  Contours of tuning in the stop-gluino
  mass plane for the (a) MSSM and the (b) NMSSM models. The vertical
  golden contours refer to the low energy values of the squark masses
  generated by universal boundary conditions at the messenger scale
  $M=300\tev$. The green line corresponds to the GMSB boundary
  conditions for the stop masses with $N=5$ messengers. The $\mu$ term
  has been fixed to 400~GeV. The yellow region corresponds to the direct 
  squark-gluino exclusion bounds from the LHC
  \cite{ATLAS-CONF-2013-047}.}
\vspace{-4mm}
   \end{center}
\end{figure}

To quantify the status of naturalness in SUSY, we start with its simplest implementation---the MSSM. 
For small $A$-terms and large $\tan\,\beta$ the value of the Higgs mass forces the stop masses to be close to $10\tev$  \cite{Giudice:2011cg,Arvanitaki:2012ps}, which implies a tuning of $10^{-3}$. 
This corresponds to the large $\tan\,\beta$ tail of the Mini-Split
family of models.

The situation improves for large $A$-terms, as the experimental value of the Higgs mass can be achieved with smaller stop masses of $\mathcal{O}(1\tev)$. In figure \ref{fig:pushup}(a), we show tuning contours in the stop-gluino mass plane. We assume universal gaugino and squark boundary conditions, and allow for extra contributions to the stop and Higgs soft masses to deviate from the GMSB prediction and fix $\mu = 400\gev$.
We also vary the $A$-terms to obtain a Higgs mass of $125\gev$; the few-GeV uncertainty in the Higgs mass translates into an overall factor of $\sim$2 in the tuning. The UV parameters used to compute tuning for this model are $a_{i} = \lbrace M_3^2, \delta m_{\tilde t}^2, \delta m_{H_u}^2, \mu^{2}, A_t \rbrace$, where $\delta m_{\tilde t}^2$ and $\delta m_{H_u}^2$ are the extra contributions to stop and Higgs soft masses, respectively.

The tuning is minimized when the stop mass is close to the gluino
mass---lighter stops are more tuned because of the strong RG attractor
effect of the gluino (eq.~(\ref{eq:gluinorunning})). At a fixed
physical stop mass, a smaller gluino mass increases the tuning because
a larger $m_{\tilde{t}}^2$ at the messenger scale is needed to
reproduce the low-energy value. This implies a larger contribution to
$m_{H_u}^2$ for most of the running.  These considerations explain the
shape of the tuning contours---minimal tuning along the $m_{\tilde
  t}\sim M_3$---which will be qualitatively the same in all of the
models studied in the rest of the paper.  The plot also shows the
current limits on squarks and gluinos from LHC searches. Taking these
into account, the tuning amounts to $\sim$0.5\% (up to a factor of 2
in either direction due to uncertainties in the Higgs mass), but is
still dominated by the large $A$-terms needed for the Higgs mass.

To further mitigate the tuning, we need to add another source for the Higgs quartic, as in the NMSSM \cite{Ellwanger:2009dp}, DMSSM \cite{Batra:2003nj}, or $\lambda$SUSY \cite{Harnik:2003rs,Barbieri:2006bg}. In figure \ref{fig:pushup}(b) we again show contours of fine tuning in the stop-gluino mass plane, taking $a_{i} = \lbrace M_3^2, \delta m_{\tilde t}^2, \delta m_{H_u}^2, \mu^{2} \rbrace$, now assuming an additional tree-level source for the quartic that fixes the Higgs mass to the measured value in a natural way.\footnote{Explicit implementations of this idea, such as the NMSSM, generically fail to achieve the right value for the Higgs mass in a completely natural way and introduce extra fine-tunings. For this reason our estimate of the tuning from direct bounds should be interpreted as a lower limit.} In this case, the LHC bounds on colored sparticles dominate the tuning which is at best $\sim$2\%, comparable to that of the MSSM. This is a shift from the pre-LHC era, when the Higgs mass bound from LEP was the primary cause of tuning in supersymmetry.

The LHC thus forces us to move beyond the minimal implementations of SUSY and look for models where the LHC bounds are less stringent. In the following sections, we consider three extensions to the (N)MSSM with GMSB boundary conditions that relax the LHC bounds by increasing the first two generation squark masses (split families), by replacing missing energy with hadronic jets (baryonic RPV), or by reducing collider limits and decoupling the gluino effects in the supersoft limit (Dirac gauginos).

\section{Split Families}
\label{sec:splitfam}

\subsection{Experimental Bounds}
\label{sec:exper-bounds-gener-u1p}

As previously discussed, the most significant contribution to the fine tuning of
the EW vacuum comes from the third generation. On the other hand, experimental
bounds from direct production at colliders are stronger on the first generation
squarks because of the larger production cross section coming from the valence
quarks of the proton, while indirect flavor bounds are dominated by $K^0$-$\bar{K}^0$
mixing and are significantly stronger on the first two generations. 
A natural way to relax the tuning is thus to keep the gluino and the third generation
squarks light while decoupling the first two generations. These
models, introduced well before the LHC
\cite{Dimopoulos:1995mi,Cohen:1996vb,Pomarol:1995xc}, are known as
split-family models.

Given that the first- and second-generation sfermions are heavy, the relevant
searches are direct production of gluinos, third-generation
sfermions, and electroweak-inos. These channels are well explored by the LHC; the
relevant limits are:
\begin{itemize}
\item $M_3 \gtrsim 1.4 \tev$ for $\tilde{g} \rightarrow t\bar
  t\chi_1^0$ and $\tilde{g}\to b\bar t \chi^+_1$, and $M_3 \gtrsim 1.3
  \tev$ for $\tilde{g} \rightarrow b\bar b \chi^0_1$
  \cite{ATLAS-CONF-2013-061}. These limits are quite robust over a wide range of
  split-family spectra, since a light higgsino acts
  as the LSP in the simplified model.  To avoid these bounds,
  the $\mu$ term has to be larger than 600--700~GeV, which leads to
  2\% tree-level tuning.
 
\item $m_{\tilde t} \gtrsim 700 \gev$ and $m_{\tilde b} \gtrsim 600 \gev$
  \cite{ATLAS-CONF-2013-024,CMS-PAS-SUS-13-004,CERN-PH-EP-2013-148}.  However, these limits
  disappear if the LSP (thus the $\mu$ term) is above $\mu \gtrsim 250
  \text{--} 300 \gev$.  Searches for cascade decays into
  a higgs \cite{CMS-PAS-SUS-13-014} and those using ISR
  \cite{Delgado:2012eu} place limits that are weaker but less dependent on the (N)LSP mass.
\item Higgsino NLSP: $\mu > 330 \gev$ from higgsino pair production promptly
  decaying to $Z$ plus MET \cite{CMS-PAS-SUS-13-006}. The limit
  might be weakened since $\text{Br}(\tilde{h}^0 \rightarrow Z \tilde{G})$ is
  model dependent \cite{Meade:2009qv,Howe:2012xe}; however, recent analyses of EWinos decaying to $h$ \cite{ATLAS-CONF-2013-093}  promise to be almost as sensitive as the those decaying into $Z$. The bound of course disappears if the Higgsino decays outside the detector.
\item Stau NLSP: for prompt decays, LEP sets a direct limit of
  $m_{\tilde \tau}> 86.6\gev$ \cite{lep-stau-bound}. Direct limits on
  promptly decaying staus from the LHC are not as strong, although
  they can still put very strong indirect bounds (e.g. $M_3>$1140~GeV
  \cite{ATLAS-CONF-2013-026}).  For long-lived staus, the LHC limit is
  $m_{\tilde \tau}> 267\gev$ \cite{ATLAS-CONF-2013-058,Chatrchyan:2013oca}.\footnote{The
    bound can get as strong as $\sim$400~GeV once other sparticles
    increase the production cross section
    \cite{ATLAS-CONF-2013-058}. Equivalently, strong indirect bounds
    can be placed on sparticles such as gluinos, stops, and EWinos that enhance the $\tilde\tau$
    production cross section. }
\end{itemize}
We focus on gauge-mediated models with multiple messengers, which are typically
less tuned; therefore we do not consider a bino NLSP, as the stau is a factor of
$\sim \sqrt{N_{\text{mess}}}$ lighter than the bino.

\subsection{Toy Model}
\label{sec:u1p-model}

The basic ingredients of split-family models are heavy sfermions of the first two generations and relatively light third-generation sfermions. 
This can be accomplished in several ways: by using a flavorful $U(1)$ \cite{Pomarol:1995xc,Dvali:1996rj,Hardy:2013uxa},
localizing the generations on different branes in an extra dimension \cite{Gabella:2007cp}, or
different sites in deconstructed versions \cite{Craig:2012hc,Craig:2011yk}, or by
gauging the approximate $SU(3)$ flavor symmetry \cite{Craig:2012di}.  Here we
are interested in a model that preserves unification and allows for a low mediation scale to minimize the tuning.
We achieve this by simply extending the usual gauge mediation mechanism with an additional gauge
group, $U(1)'$, under which only the first two generations are charged. By choosing the charge to be $B$$-$$L$, 
the $U(1)'$ is automatically anomaly free, can be implemented at low scales and is 
compatible with unification. Further details of the model can be found in appendix \ref{sec:u1p-appendix}.

The resulting sfermion masses consist of a universal gauge-mediated contribution
and an additional contribution only to the first two generations:
\begin{equation}
  m_{\phi_{i}}^2 \sim 2\sum\limits_{a}
  C_a(i)\left(\frac{\alpha_{a}}{4\pi} \frac{F}{m_D}\right)^2 +
  (\delta_{1 i}+\delta_{2
    i})(B_i-L_i)^2\left(\frac{\alpha_{\phi}}{4\pi}\frac{F}{m_N}\right)^2,\\
\end{equation}
where $m_N$ and $m_D$ are $U(1)'$ and the usual $\bold{5}+\bold{\bar{5}}$ gauge
messenger masses, respectively, and $i$ is the generation index. We absorb the charge of the $U(1)'$ gauge messenger into the definition of the $U(1)'$ gauge coupling $\alpha_{\phi}$. 
For simplicity we took the same SUSY breaking parameter $F$ for the two contributions, so that 
the ratio $\alpha_\phi m_D/(\alpha_a m_N) $ controls the mass splitting between the first two and the third generations.

To generate the mixing between the light-flavor and heavy-flavor quarks
(e.g. $\mathcal{W} \supset H_d Q_1 D_3$ which is forbidden by the $U(1)'$
symmetry), we use another $\bold{5}+\bold{\bar{5}}$ pair, $D''$ and $
\bar{D}''$, which does not couple to SUSY breaking,\footnote{When the
  pair $D''$ and $\bar{D}''$ is coupled directly to SUSY breaking it
  generates too large tree-level SUSY breaking contributions.} with superpotential
couplings
\begin{equation}
  \label{eq:u1p-superpotential-mixing}
  \mathcal{W} \supset H_dQ_iD'' + \Phi \bar{D}'' D_3 + m_{D''}D''
  \bar{D}'' + V(\Phi)_{U(1)'\text{-breaking}} \quad\quad (i = 1,2).
\end{equation}
$\Phi$ is a singlet under SM gauge groups and breaks the $U(1)'$ symmetry when
it acquires a vev. Then the low-energy effective theory below $m_{D''}$ contains the
necessary $H_dQ_{i}D_3$ terms.

Split family models generically have non-trivial sflavor structure which can
have experimental signatures, e.g. in the $K^0$-$\bar{K}^0$ system. In our
model, the contributions to flavor observables are within theoretical
errors of the SM prediction (see appendix \ref{sec:u1p-appendix}).

\subsection{Fine Tuning}
\label{sec:u1p-fine-tuning}
\begin{figure}[t]
  \centering
  \includegraphics[width=0.8\textwidth]{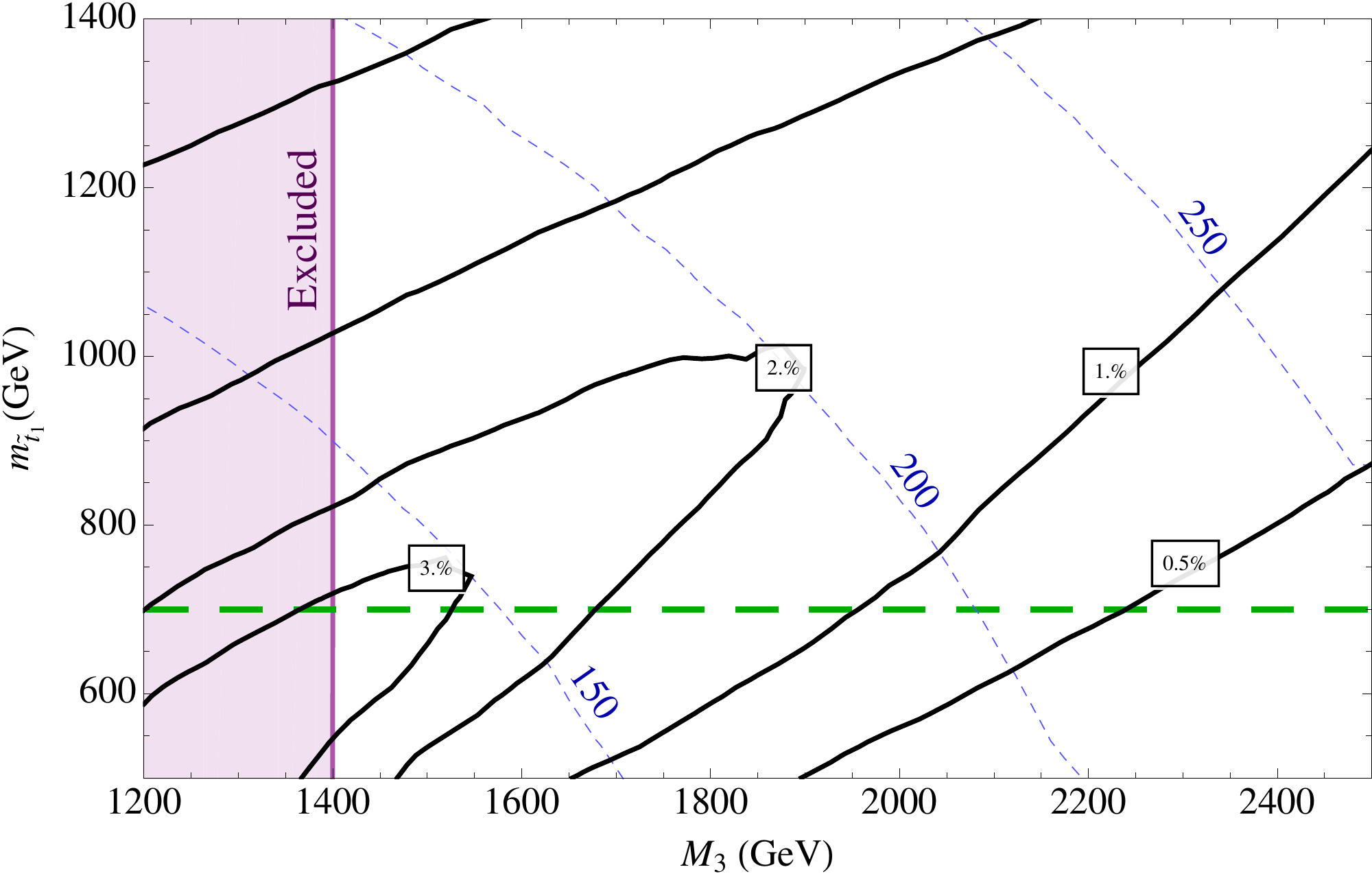}
  \caption{Fine tuning in the split-family model. In this plot, we use 4 pairs
    of $(5+\bar{5})$ messengers and fix $F/m_N^2 = 0.8$, $\alpha_{\phi} = 0.1$,
    $\tan \beta = 5$, and the first two generations at $\sim 8.5 \tev$. These
    parameters determine the mediation scale $m_D \sim 10^7 \gev$. Additional
    soft mass contributions to $m_{H_u}^2$ and $m_{H_d}^{2}$ are chosen such that
    $\mu \sim 300 \gev$, although the fine tuning is relatively insensitive to
    the choice of $\mu$ for $\mu \lesssim 400 \gev$. The solid black lines are
    FT contours, and the dashed blue lines are the stau mass contours. The
    shaded region corresponds to the gluino bound, and the dashed green line
    indicates the limit from direct stop production, which applies for $\mu
    \lesssim 250\gev$. }
  \label{fig:u1p-ft}
\end{figure}
In figure \ref{fig:u1p-ft}, we present the fine-tuning with respect to
the fundamental parameters: $F$, which characterizes the overall soft
mass scale, and $m_N$, which characterizes the splitting between the
generations. As discussed in section \ref{sec:QN}, we are agnostic to
the solution of the $\mu$-$B\mu$ problem, and therefore additional
contributions to $m_{H_u}^2$, $m_{H_d}^{2}$, and $m_{\tilde{t}}^2$ are
treated as free parameters and included in the FT calculation.

Split-family models can be less tuned than the (N)MSSM: squark pair
production and associated squark-gluino production are effectively
turned off if light-flavor squarks are decoupled, so the stop can be
lighter than the degenerate squark-gluino bound of $\sim 1.6 \tev$
\cite{ATLAS-CONF-2013-047}. However, the improvement is limited
because the gluino contribution to fine tuning is not diminished in
split-family models \footnote{For another analysis of tuning in
  split families, see \cite{Hardy:2013ywa}.}. In particular, the stop
mass is still pulled up by the gluino mass; while there is an extra
negative contribution to the stop from the heavy first and second
generation squarks, using this effect to lower the stop mass is a
cancellation between two independent contributions and does not
improve the tuning.

We can see this effect when comparing to the fine tuning in the NMSSM in
figure \ref{fig:pushup}(b): for a given level of tuning, the maximum
gluino mass allowed is comparable between the NMSSM and the split-family
model, while the maximum stop mass allowed is typically a few hundred
GeV lower in the split families model. Therefore bounds from
direct stop pair production, when applicable, are probing the
same level of tuning as the gluino bounds.

For $\mu \gtrsim 250\gev$, bounds from direct stop pair production no longer
apply, but in this case, bounds from pair production of staus provide a
complementary probe. Limits on promptly decaying staus are relatively weak, but
light staus, especially from cascade decays, are long-lived: the soft masses of the first two generations determine $F \gtrsim \left(\frac{4\pi}{\alpha_{\phi}}
  m_{1,2}\right)^2 \sim 10^{12}\gev^2$, which in term determines $c\tau$ of the
stau NLSP to be $\gtrsim 1$ m ($10^{-3}$ m) for stau mass of 100 GeV (300
GeV). 

While a careful study of the stau bound is beyond the scope of this work, it may
actually be the most relevant constraint for the naturalness in this class of
models, as can be seen in figure~\ref{fig:u1p-ft}.  Taking into account all
experimental limits, we conclude that split-family models are tuned to at best 3\%.

\section{Baryonic RPV}
\label{sec:rpv}

An experimentally-driven way to reduce limits on sparticles is to
eliminate the tell-tale missing energy signature of a stable LSP by
allowing some degree of $R$-parity violation (RPV).  The presence of
both $B$- and $L$-violating operators would lead to proton decay, and
current limits on proton lifetimes set stringent bounds of $10^{-14}
\mathrm{-}10^{-27}$ on products of $B$ and $L$-violating couplings. In
addition, LHC
limits on $L$-violating operators are comparable to or stronger than
those on the R-parity conserving ones (see
e.g. \cite{Barbier:2004ez,ref:cmsrpv}).  Given these strong
contraints, our best hope of finding natural models in the RPV context
lies with models that conserve lepton number and only violate baryon
number via the operator $\mathcal{W}\supset\half \lambda''_{ijk}U^c_i
D^c_j D^c_k$. There has been a renewed interest in baryonic RPV
(bRPV), both in proposed and experimental searches and model-building
(see e.g.  \cite{Evans:2012bf,
  Han:2012cu,Allanach:2012vj,Brust:2012uf,Chatrchyan:2012uxa,ATLAS:2012dp}
and \cite{Bhattacherjee:2013gr,Franceschini:2013ne,DiLuzio:2013ysa}).

\subsection{Constraints on bRPV couplings}

\subsubsection*{Collider Limits}

The LEP experiment places limits on charginos $m_{\tilde\chi^{\pm}} >
103\gev$ independent of the presence of RPV, as well as on stops and
sbottoms decaying to jets $m_{\tilde b,\,\tilde t} \gtrsim 100\gev$
\cite{Heister:2002jc}. CMS searches for gluinos decaying to
3 light-flavor jets place limits of $M_3\gtrsim 666\gev$ with data from the 
$7\tev$ run and a recent ATLAS search for gluinos decaying to decaying to
$t \tilde t^*$ with $\tilde t^* \rightarrow bs$ sets a gluino limit of
$860\gev$ for stop masses up to $1\tev$ \cite{Chatrchyan:2012uxa,ATLAS:2012dp, ATLAS-CONF-2013-007}. Even more recently  \cite{ATLAS-CONF-2013-091}, ATLAS searches for pair-produced
gluinos decaying to 6 or 10 jets set limits between
$\sim 650\gev$ and $\sim 950\gev$ depending on the final state
quarks; including the associated squark-gluino production in the cross-section could raise the limits to
over a TeV. While these searches
place important constraints, they are preliminary and sensitive to the
sparticle spectrum and value of $\lambda''_{ijk}$ couplings. Other
searches which are used for R-parity conserving SUSY can be recast to
place limits on hadronic RPV spectra; for instance multijet and
same-sign lepton searches were used in \cite{Asano:2012gj} to set
limits of $800$-$900\gev$ on gluinos and first- and second-generation
squarks with just the $7 \tev$ dataset.

\subsubsection*{Low-energy Limits }

While LHC experiments are only recently beginning to search for RPV
decays of sparticles, there is a multitude of indirect limits on
$B$-violating couplings from rare decays and low-energy experiments
(for a review see \cite{Barbier:2004ez}):
\begin{itemize}
\item The missing energy signature at colliders can be restored if 
  the LSP decays out of the detector, or the NLSP has a sizable
  decay width to a light gravitino.  The former requires, for
  neutralino 3-body decays,
  \begin{equation}
    \label{eq:6}
    \lambda''_{ijk}>7\times 10^{-6}\left(\frac{m_{\tilde\chi
        0}}{125\gev}\right)^{-5/2}\left(\frac{m_{\tilde q
      }}{600\gev}\right)^{2}\left(\frac{1
      \,\mathrm{m}}{c\tau}\right)^{1/2} \left(\frac {\zeta}{0.1}\right)^{-1},
  \end{equation}
  where $\zeta$ is the neutralino-squark-quark coupling.
   Two-body right-handed squark
  decays give
  \begin{equation}
    \label{eq:1}
    \lambda''_{ijk}>10^{-9}\left(\frac{m_{\tilde q
      }}{600\gev}\right)^{-1}\left(\frac{1
      \,\mathrm{m}}{c\tau}\right)^{1/2},
  \end{equation}
  while left-handed squark decays are further suppressed by left-right
  mixing; we will focus on a neutralino NSLP because of naturalness. We
  discuss limits from decay into  the gravitino in the following section.
\item Violation of baryon number washes out the results of
  baryogenesis unless $\lambda''_{ijk}<10^{-7}$; if the reheat temperature is low these bounds can be avoided \cite{Dimopoulos:1987rk,Barbieri:1985ty}. Recently, it has been pointed out that  the baryogenesis problem in RPV models  can be also addressed without invoking a low reheat temperature \cite{Cui:2012jh}.
\item Low-energy measurements such as neutron-antineutron oscillations
  and di-nucleon decay; the relevant limit for our discussion is
  $\lambda_{1jk}'' \lesssim 10^{-5} \left(\frac{m_{\tilde
        q}}{600\gev}\right)^2\left(\frac{M_3}{600\gev}\right)
  $ \cite{Goity:1994dq}.
\end{itemize}
Some couplings, e.g. $\lambda_{323}''$, have no direct experimental
constraints; however, a concrete model of RPV couplings connects these
to more constrained couplings. In order to avoid flavor constraints,
generic models will relate the UDD R-parity violating couplings to
powers of the corresponding yukawa couplings.  In the model we
consider below, for example, the couplings that provide the strongest
constraints are the first generation of up-type couplings
$\lambda''_{1jk}$. The relevant limits are shown in figure
\ref{fig:rpvlambda}, and more details can be found in
appendix~\ref{app:rpv}.
\begin{figure}[t]
\begin{center}
   \includegraphics[trim = 0mm 0mm 0mm 0mm, clip, width
      = 0.65\textwidth]{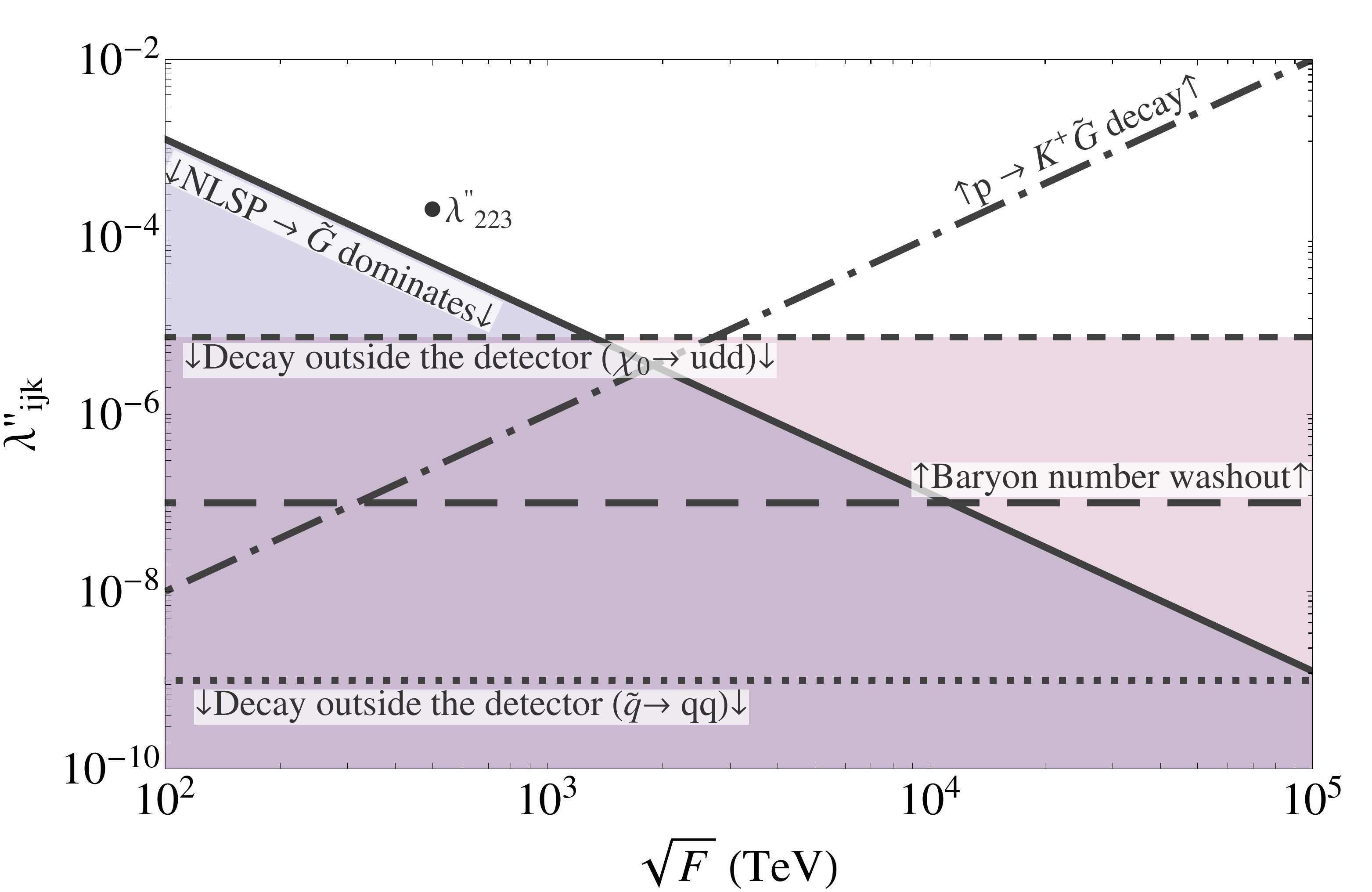}
\vspace{-2mm}
\caption{Limits on RPV UDD coupling $\lambda''_{ijk}$ and scale of
  SUSY breaking $\sqrt{F}$ when one coupling $\lambda''_{\text{max}}$
  dominates. The region below the dashed line is excluded as the NLSP
  would decay outside the detector, restoring the MET signature (if
  the only available decay is 3-body).  For low $\sqrt{F}$, the decay
  rate to the gravitino can exceed the RPV decay rate (solid line),
  again leading to MET. Baryogenesis and proton decay bounds are
  indicated by the thick-dashed and the dot-dashed lines,
  respectively. There may be up to $\mathcal O (10)$ uncertainty in
  some of the bounds presented. Without extra assumptions, the entire
  parameter space is excluded; we assume low-scale baryogenesis and an
  additional SUSY-breaking sector to raise the gravitino mass and
  avoid proton decay. Since the NLSP is below the top threshold, the
  LHC phenomenology of the toy model is controlled by the
  2nd generation coupling $\lambda''_{2jk}$, and we use a low
  SUSY-breaking scale of $\sqrt{F}=500\tev$, as indicated by the gray
  point. } \label{fig:rpvlambda} \vspace{-2mm}
   \end{center}
\end{figure}

\subsubsection*{The Gravitino}

In theories with low-scale SUSY breaking (as favored by
naturalness), there is a light gravitino with mass $m_{\tilde
  G}\gtrsim\kev\times\frac{F}{(10^3\tev)^2}$. In the presence of RPV,
proton decay via $p\rightarrow K^+ \tilde G$ results in a strong
constraint on UDD couplings, which is particularly stringent on
light-flavor RPV:
\begin{equation}
  \lambda''_{ijk}\lesssim 10^{-6}
  \frac{F}{(10^3\tev)^2}\left(\frac{m_{\tilde
        s_R}}{600\gev}\right)^3,\quad \mathrm{and} \quad \lambda''_{112}\lesssim10^{-12}
\frac{F}{(10^3\tev)^2}\left(\frac{m_{\tilde s_R}}{600\gev}\right)^2.
\label{eq:ptogdecay}
\end{equation}
With all RPV couplings less than $\mathcal{O}(10^{-6})$, sparticle RPV
decays compete with decays to the light gravitino. In case the
gravitino decay is order one---or even $10\%$---the missing energy
signature of $R$-parity conserving SUSY is effectively
restored.\footnote {The relevant searches have efficiencies that are
  relatively insensitive to the superpartner masses (e.g. photon and
  missing energy \cite{Chatrchyan:2012jwg}), so the scale of SUSY
  breaking must be such that the branching ratio to gravitino times the
  production cross-section is less than the experimental limit on
  $\sigma$ (a few fb).} This constraint restricts the scale of
SUSY breaking through the decay rate to the gravitino,
\begin{align}
  \label{eq:3}
 \lambda_{max}'' &\gtrsim 4\times 10^{-5}\,\left(\frac{F}{ 10^6\tev^2}\right)^{-1}\left(\frac{m_{\tilde{q}}}{600\gev}\right)^2 \left(\frac{\zeta}{0.1}\right)^{-1}\left(\frac{\mathrm{Br}(\tilde
      \chi_0 \rightarrow\tilde G X)}{10\%} \right)^{-1/2}.
\end{align}
where $\lambda_{max}''$ is the coupling that dominates the neutralino RPV
decay.  Given the strong constraints from equations
\eqref{eq:ptogdecay} and \eqref{eq:3}, it is useful to consider a
gravitino mass above the proton mass. A gravitino heavier than $1\gev$
can decay through baryon number violation to three quarks which then
hadronize, $\tilde G~\rightarrow~u_id_jd_k$, with a lifetime
\begin{align}
\tau_{\tilde G}&\simeq 10^{7}\mathrm{s}
\left(\frac {10\gev}{m_{\tilde
      G}}\right)^9\left(\frac{m_{\tilde
      q}}{600\gev}\right)^4\left(\frac{\sqrt{F}}{100\tev}\right)^4\left(\frac{10^{-4}}{\lambda''_{2jk}}\right)^2.
\end{align}
Note that here $F$ of our sector can be less than $m_{\tilde G} M_{\mathrm
  {pl}}$, so the decay is relatively faster. The lifetime ranges from
as short as $10^{-5}$~s for $\lambda''_{2jk}=0.1$ and a $20\gev$
gravitino, to longer than $10^{19}$~s for a $2\gev$ gravitino with $1200\gev$
squarks and $F$ of $(10^3\tev)^2$.

Even without committing to a particular hierarchy of RPV couplings, it
is impossible to satisfy all the constraints discussed above (as shown
in figure~\ref{fig:rpvlambda}). We choose to give up the limit from
baryogenesis. We also introduce extra hidden sector(s) to avoid proton
decay limits while keeping a low mediation scale; these sectors have
SUSY breaking which lifts the gravitino mass up to $1$--$15\gev$ while
the $F$ corresponding to the SUSY breaking scale in our sector is
lower.  To avoid the limits discussed above, we need the gravitino
with $m_{\tilde G}> 1\gev$, but not too heavy: flavor problems from
gravity mediation arise for $m_{\tilde G}/m_{\tilde q} \gtrsim 0.02
\times \sqrt{m_{\tilde q}/\tev}$,
while for higher gravitino mass, the gravity-mediated contributions
dominate the tuning due to the very large amount of running.

\subsection{Toy Model}
\label{rpv_model}
Given the constraints discussed above, we impose the following
requirements on a model of R-parity violation: gauge coupling
unification; no lepton-number violation; low scale of SUSY breaking to
reduce fine-tuning; gauge mediation to avoid flavor issues;
and a hierarchy of couplings between the very
constrained first-generation $\lambda''_{1jk}$  couplings and the less
constrained $\lambda''_{2jk}$, $\lambda''_{3jk}$ couplings. 

We use an orbifold GUT model with $SU(5)$ and matter in the bulk, and
$SU(3)\times SU(2)\times U(1)$, the Higgses, and B-violation on the IR
brane. Doublet-triplet splitting solutions in these models allow us to
achieve baryon breaking number without lepton number while preserving
the prediction of unification. This results in a predictive pattern of
RPV couplings,
\begin{equation}
\lambda''_{ijk}\simeq\frac{\vev{\Phi}^2}{M_{D}^2}\sqrt{y^u_i} \left(\frac{y^d_j}{\sqrt{y^u_j }} \right) \left(\frac{y^d_k}{\sqrt{y^u_k }} \right),
\label{eq:coupl}
\end{equation}
where $M_D$ is the mass of a heavy $D$-type field and the vev of
$\Phi$ breaks baryon number. For more details of the model, see
appendix~\ref{app:rpv}. 

The pattern of RPV couplings results in the limits of
\begin{equation}
  \label{eq:5}
  \lambda''_{1jk} \lesssim 10^{-5},\quad \lambda''_{2jk} \lesssim
2\times 10^{-4}, \quad \mathrm{and}\quad \lambda''_{3jk} \lesssim 3\times 10^{-3},
\end{equation}
for squark and gluino masses of $600 \gev$ or higher. In order to minimize the tuning,
we maximize the value of ${\vev{\Phi}^2}/{M_{D}^2}$---thus the RPV
couplings---to minimize the branching ratio of the NLSP to the goldstino and achieve the lowest SUSY breaking scale possible.

\subsection{Fine Tuning}
\label{rpv_tuning}

\begin{figure}[t]
\begin{center}
   \includegraphics[trim = 0mm 0mm 0mm 0mm, clip, width
      = 0.8\textwidth]{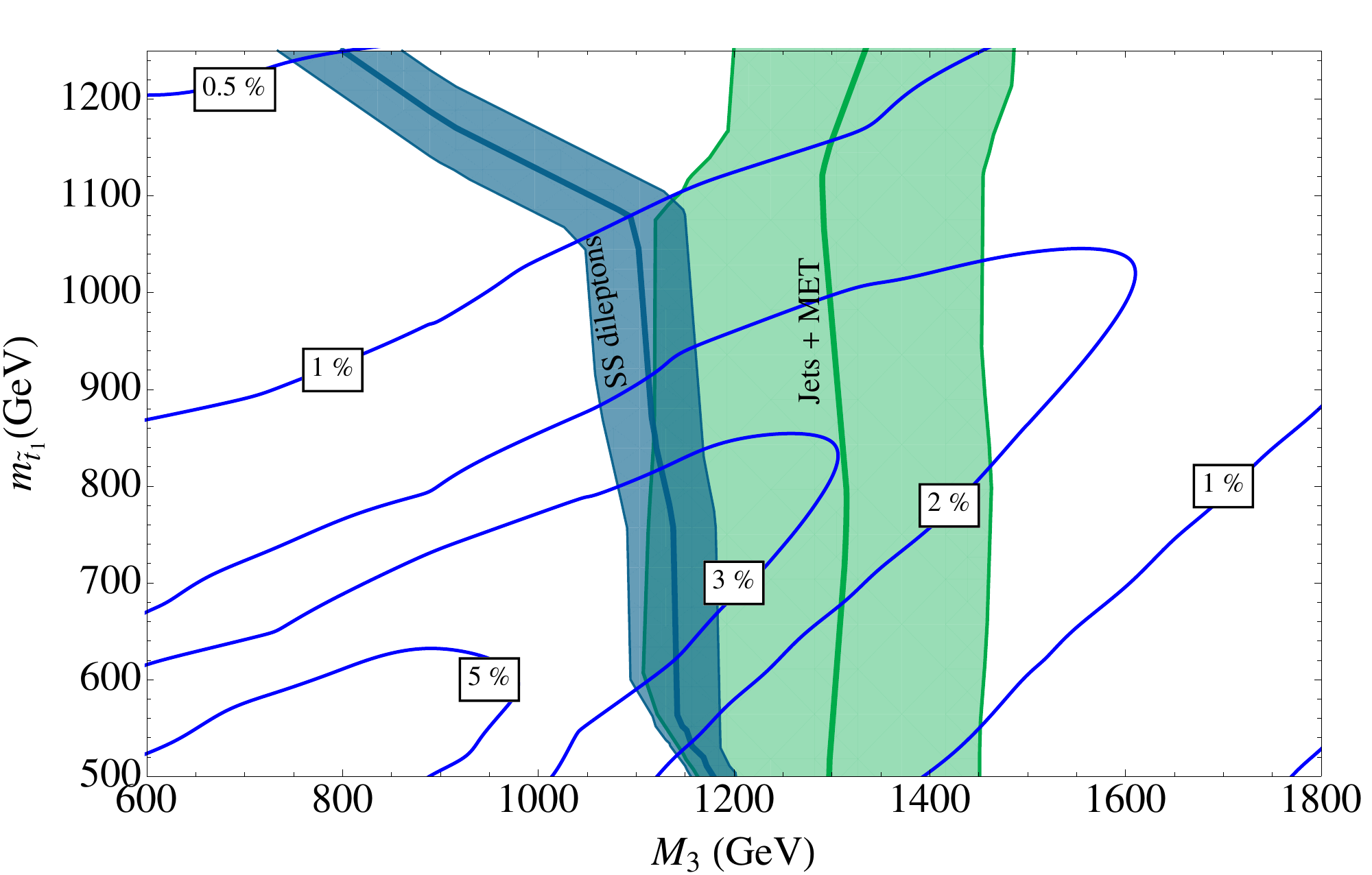}
      \caption{Tuning in RPV with standard gauge-mediated gaugino
        masses and $\mu=125 \gev$. The blue band is the recast CMS same-sign dilepton limit ($10\,
        \text{fb}^{-1}$) \cite{CMS-PAS-SUS-13-013}, and the green band
        is the recast ATLAS 7-10 jets and MET limit ($20 \,\fb^{-1}$)
        \cite{ATLAS-CONF-2013-054}. The width of the band is the
        uncertainty in our estimate; see appendix \ref{app:rpv} for
        details.  The NLSP is a neutralino which decays through bRPV, $\chi_0\rightarrow
        UDD$.} \label{fig:naturalrpv}
   \end{center}
\end{figure}

The most stringent limits come from multiple jet and MET searches
\cite{ATLAS-CONF-2013-054,Chatrchyan:2013lya} as well as leptonic
searches \cite{Chatrchyan:2012paa,ATLAS-CONF-2013-007}. These are
shown in figure~\ref{fig:naturalrpv}; for details of event simulations,
analysis, and validation, as well as comparison to other re-castings,
see appendix~\ref{app:rpv}. We calculate fine tuning against UV
parameters as follows,
\begin{equation}
\text{FT}_\text{total} =  \text{FT}_{m_{\tilde t}^2}\left[\delta m_{H_u}^2, \delta m_{\tilde t}^2,\Lambda^2\right]\times \text{FT}_{m_h^2}\left[\mu, \delta m_{H_u}^2,\delta m_{\tilde t}^2,\Lambda^2\right].
\end{equation}

In generic spectra, leptons and missing energy resulting from
electroweakino and top decays are relatively common in cascades. The
NLSP is a combination of bino and higgsino and decays predominantly
through $\lambda''_{2jk}$ to three jets, possibly including a
$b$-jet. We fix the $\mu$ term at $125\gev$, and the benefit is
two-fold: light higgsinos are beneficial for naturalness, and an NLSP
below the top mass reduces the number of leptons in cascades, as
otherwise the NLSP would decay dominantly to a top through the larger
third-generation $\lambda''_{3jk}$ couplings. The limits on the gluino
and squark masses are much reduced compared to R-parity-conserving
scenaria; however, the tuning of the models is still at the few percent
level as shown in figure~\ref{fig:naturalrpv}.

There are two avenues to try to further reduce the bounds. One is to
remove leptons and neutrinos completely from the decay chain, but this
generally leads to increased tuning. For example, we found that a
spectrum with a heavy wino has comparable limits from same-sign (SS)
leptons to figure \ref{fig:naturalrpv}, but tuning increases by a
factor of 2 to 3 due to the larger left-handed stop mass and the
addition of an extra parameter. The leptons in this case arise from
e.g. $\tilde g \rightarrow \tilde b_R b \rightarrow \tilde \chi^- t b$
decays as well as transitions between the remaining electroweakinos
through $W$s and $Z$s. The latter transitions can come from off-shell
decays of the charged higgsinos to the neutral higgsino, which still
dominate over direct RPV decays due to a relatively small RPV
coupling. Another possibility is a very light spectrum where decays
through on-shell tops are kinematically forbidden, for example
higgsinos at $150\gev$, squarks at $300\gev$ and gluinos at
$450\gev$. However, this scenario is ruled out by ATLAS multijet
searches for RPV gluinos
\cite{ATLAS:2012dp,ATLAS-CONF-2013-091}.\footnote{We thank Prashant
  Saraswat for confirming that such light spectra are excluded.}

Another approach is to reduce the very large colored production
cross-sections by decoupling the first two generation squarks as in
split families. However, a light gluino is still required by
naturalness; such a split RPV spectrum is covered by an ATLAS search
for $\tilde g\rightarrow \tilde t\overline t$, with $\tilde t \rightarrow j\,j$ as
mentioned previously \cite{ATLAS-CONF-2013-007}, resulting in a limit
of $860 \gev$ on the gluino mass from SS leptons, assuming the
only leptons arise from a top on either side of the decay
chain. Introducing new sources of leptons in the cascade will increase
the bound. In particular, a light higgsino, required for naturalness,
would further increase this limit by adding more tops to the cascade:
the stops will decay $\tilde t\rightarrow \tilde H^0 t$ with a large
rate due to the large yukawa coupling, increasing the incidence of SS
leptons by nearly a factor of $4$ which would raise the limit to
$1050\gev$ for gluino production alone. In addition, by assumption the
bino and wino are decoupled, which further increases the tuning, and
the mechanism to split the families also introduces more tuning
c.f.~section~\ref{sec:splitfam}.

One alternative we have not considered so far is a stop NLSP that
decays to two jets via a UDD coupling. Here we can apply the bound of
\cite{ATLAS-CONF-2013-007} at face value, so a $860\gev$ gluino is
allowed; the optimal tuning in our model would be about $5\%$ for a
$600\gev$ stop. However, this does not take into account the remainder
of the tunings required to achieve this spectrum. First, the
cross-section used to set the above limit assumes only gluino
pair-production; splitting the families to tends to further increase
the tuning, or alternatively, the increased production cross-section
from the presence of light squarks leads to a limit of $1100\gev$ on
the average gluino-squark mass. Also, as discussed above, it is
necessary to raise the higgsinos to avoid additional tops in the
cascade. Fixing the $\mu$ term at $400\gev$ or higher to reduce the
branching fraction of $\tilde t\rightarrow \tilde H^0 t$ gives $5\%$
tuning from the $\mu$ term; alternatively, lowering the stop mass
against the gluino mass introduces additional tuning of at least a
factor of $(600/m_{\tilde t})^2$. These values do not take into
account adding these tunings in quadrature with other sources, such as
new parameters to decouple the wino and bino or the sleptons. In fact,
including only the tunings due to the $\mu$ term and to splitting the
families, a stop at $600\gev$ and higgsino at $400\gev$ result in
FT $\sim 3.5\%$, at the same level as the neutralino NLSP case
we discuss above.

\section{Dirac Gauginos}
\label{sec:dirac}
One possible way to extend the MSSM is to impose an exact $R$-symmetry, and---as Majorana gaugino masses are now forbidden---to postulate that the gauginos $\lambda_i$ acquire Dirac mass terms with the fermions of extra adjoint superfields $A_i$ \cite{Fayet:1978qc,Polchinski:1982an,Hall:1990hq,Fox:2002bu,Nelson:2002ca,Antoniadis:2006uj,Amigo:2008rc}.  $R$-symmetric Dirac gaugino models have been heralded as beneficial for fine-tuning of the Higgs mass \cite{Kribs:2012gx,Fox:2002bu}.  The stops still contribute to the Higgs mass and quartic, but the gluino can be naturally heavy without ``pulling up'' the squark masses, removing the dominant contribution to fine-tuning given the strong LHC bound on the gluino mass \cite{Arvanitaki:2012ps}. Bounds on direct squark pair production are milder, because $t$-channel gluino exchange decouples faster for a heavy Dirac gluino than for a heavy Majorana gluino \cite{Heikinheimo:2011fk}.  Finally, flavor bounds are significantly weakened in the absence of Majorana gaugino masses \cite{Kribs:2007ac}.  

From the outset unification is difficult to achieve, as the $SU(2)$ and $SU(3)$ adjoint fields dramatically change the running of the gauge couplings (see \cite{Benakli:2010gi} for a one-loop analysis). In particular, at scales above the mass of the chiral octet $A_3$ the strong gauge coupling is no longer asymptotically free, and at two-loop order runs stronger according to:
\begin{align}
\partial_t \alpha_3 = 0 \, \alpha_3^2 + \frac{136}{(4\pi)^2} \alpha_3^3 + \dots.\label{eq:alpha3running}
\end{align}
To prevent a Landau pole before the scale $10^{16} \gev$ (which would lead to unacceptably rapid proton decay) any pair of colored states needs to be heavier than $\sim 10^{12} \gev$.  The minimal representation of $SU(5)$ containing the $A_i$ is the $\mathbf{24}$, which also has ``bachelor" fields $(3,2,-5/6)+(\bar{3},2,5/6)$.  The mass of these bachelor fields has to be above $\sim 10^{14}\gev$ to avoid a Landau pole before $10^{16} \gev$. It is also possible to achieve perturbative unification with additional light electroweak states in incomplete multiplets. In either case, the quantitative \emph{prediction} of gauge coupling unification in the MSSM is lost, since additional states need to be added at intermediate scales.

The above-mentioned positive aspects of Dirac gaugino models stem from the ``supersoft" operator
\begin{align}
\mathcal{W} &\supset  \sqrt{2} m_{Di} \theta^{\alpha}  W^a_{i\alpha} A^a_i \qquad \Rightarrow \qquad \mathcal{L} \supset -  m_{Di} \lambda^a_i \tilde{A}^a_i - \sqrt{2}m_{Di} \left(A_i^a + A_i^{a\dagger} \right)D_i \label{eq:softmass1},
\end{align}
which contains Dirac masses for the gauginos labeled by $i = 1,2,3$.  After integrating out the auxiliary $D$-terms, it also gives rise to mass terms for the real component of the scalar adjoints, $\Re(A^a_i)$, and a tri-scalar coupling of $\Re(A_i^a)$ with the MSSM sfermions.
In addition, the scalar adjoints can also have SUSY-breaking, $R$-symmetry-preserving mass terms:
\begin{align}
\mathcal{L} \supset  m_{A_i}^2 A_i^{a\dagger} A^a_i+B_{A_i} \left(A^a_i A^a_i +\text{h.c.}\right). \label{eq:softmass2}
\end{align}
From eqs.~(\ref{eq:softmass1}) and (\ref{eq:softmass2}), the real and imaginary components of the adjoint scalars have masses
\begin{align}
m^2_{\Re (A_i)} = 4m_{Di}^2 + m_{A_i}^2 + B_{A_i} \qquad m^2_{\Im(A_i)} =  m_{A_i}^2 - B_{A_i}.\label{eq:masseigenstates}
\end{align}
The sfermion masses do not receive any log-divergent one-loop corrections proportional to $m_{Di}$, but do get finite threshold contributions due to the mass splittings of the Dirac gauginos and their real scalar adjoints \cite{Fox:2002bu}:
\begin{align}
\Delta_\text{finite}m^2_{\tilde{f}} = \sum_i \frac{C_i(f) \alpha_i m_{Di}^2}{\pi} \log \frac{m_{\Re(A_i)}^2}{m_{Di}^2} \label{eq:sfermionmasses}.
\end{align}

LHC bounds on Dirac gaugino models can be relatively mild if Dirac gluino pair production and associated production are not kinematically accessible. The cross-section for direct squark pair production is reduced due to the suppression of the $t$-channel (Dirac) gluino exchange diagram, leading to reduced limits for colored sparticles at the LHC \cite{Kribs:2012gx,Heikinheimo:2011fk}.  In a squark-LSP simplified model with decoupled gluinos, CMS and ATLAS place bounds of $m_{\tilde{q}} \gtrsim 800 \gev$ if $m_{\text{LSP}} \lesssim 300\gev$ with $\sim 20\,\text{fb}^{-1}$ of data \cite{CMS-PAS-SUS-13-012,ATLAS-CONF-2013-047}; CMS has an earlier search using razor variables which excludes $m_{\tilde{q}}\lesssim 600 \gev$ for heavier LSP masses as well \cite{CMS-PAS-SUS-12-005}. Very compressed spectra with low masses may still be allowed, but since they are not generic in Dirac gaugino models we do not consider these here. The NLSP is typically the higgsino or a right-handed slepton; the gravitino may also be relevant for collider limits depending on the scale of SUSY breaking. If the NLSP is a higgsino and decays to a gravitino via a $Z$ boson inside the detector, the higgsino mass is constrained to be $\mu \gtrsim 360\gev$ \cite{CMS-PAS-SUS-12-022}.  Direct searches for sleptons \cite{CMS-PAS-SUS-13-006} can be important as well depending on their masses and the decay width to the gravitino, although they are less relevant for naturalness of Dirac gaugino models.

In gauge-mediated models of Dirac gauginos
\cite{Benakli:2008pg,Carpenter:2010as,Fox:2002bu,Benakli:2012cy}, it
is assumed that the UV soft mass terms in eqs.~(\ref{eq:softmass1})
and (\ref{eq:softmass2}) arise from a perturbative coupling of the
adjoint superfields with sets of messengers $X_j, X^c_j$ in the
superpotential $\mathcal{W} \supset M_{jk} X_j X_k + \sum_i y_{ijk}
X^c_j A_i X_k$. If the messengers are charged under a $U(1)'$ that
acquires a $D$-term expectation value, $m_{D_i}$, $m_{A_i}^2$, and
$B_{A_i}$ are generated at one loop
\cite{Carpenter:2010as,Benakli:2008pg}.  In a minimal model with one
pair of messengers, $m_{A_i}^2$ is suppressed at leading order in
$D^2/M^2$, which causes the real scalar adjoint to be
tachyonic. However, it is possible to give positive masses to both the
real and imaginary scalar adjoints with multiple sets of messengers
provided there is sufficient mixing between them
\cite{Benakli:2008pg}, or with additional ($R$-symmetric) $F$-terms
\cite{Benakli:2010gi}. Mediation schemes with both non-tachyonic
adjoints thus naturally generate the mass terms $m_{A_i}^2$ and
$B_{A_i}$ a loop factor higher than $m_{Di}^2$.

\begin{figure}[t]
\begin{minipage}[t]{0.485\textwidth}
   \vspace{1pt} \subfigure[]{\includegraphics[width=\textwidth]{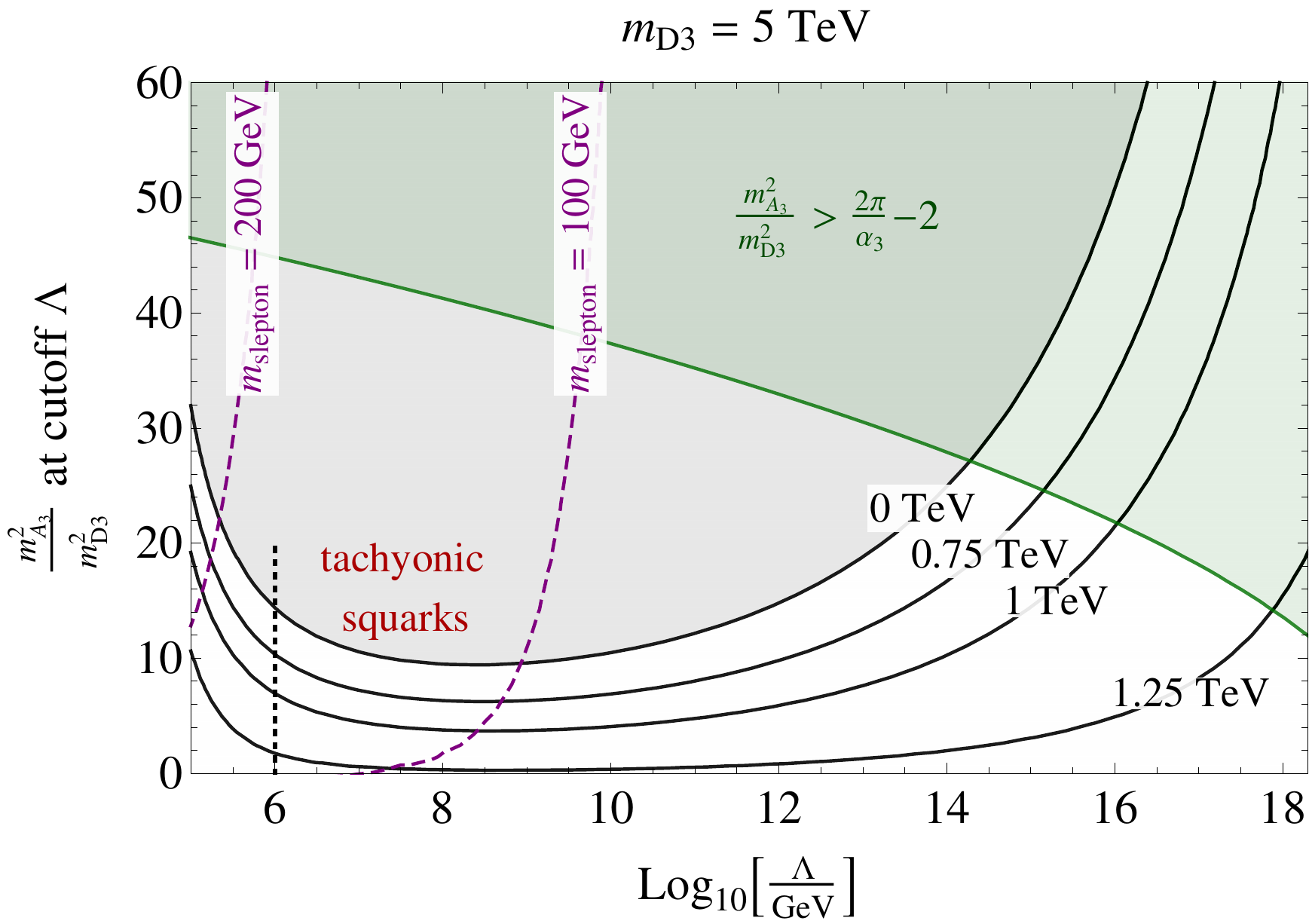}}
\end{minipage}
  \hfill
\begin{minipage}[t]{0.495\textwidth}
   \vspace{0pt} \subfigure[]{\includegraphics[width=\textwidth]{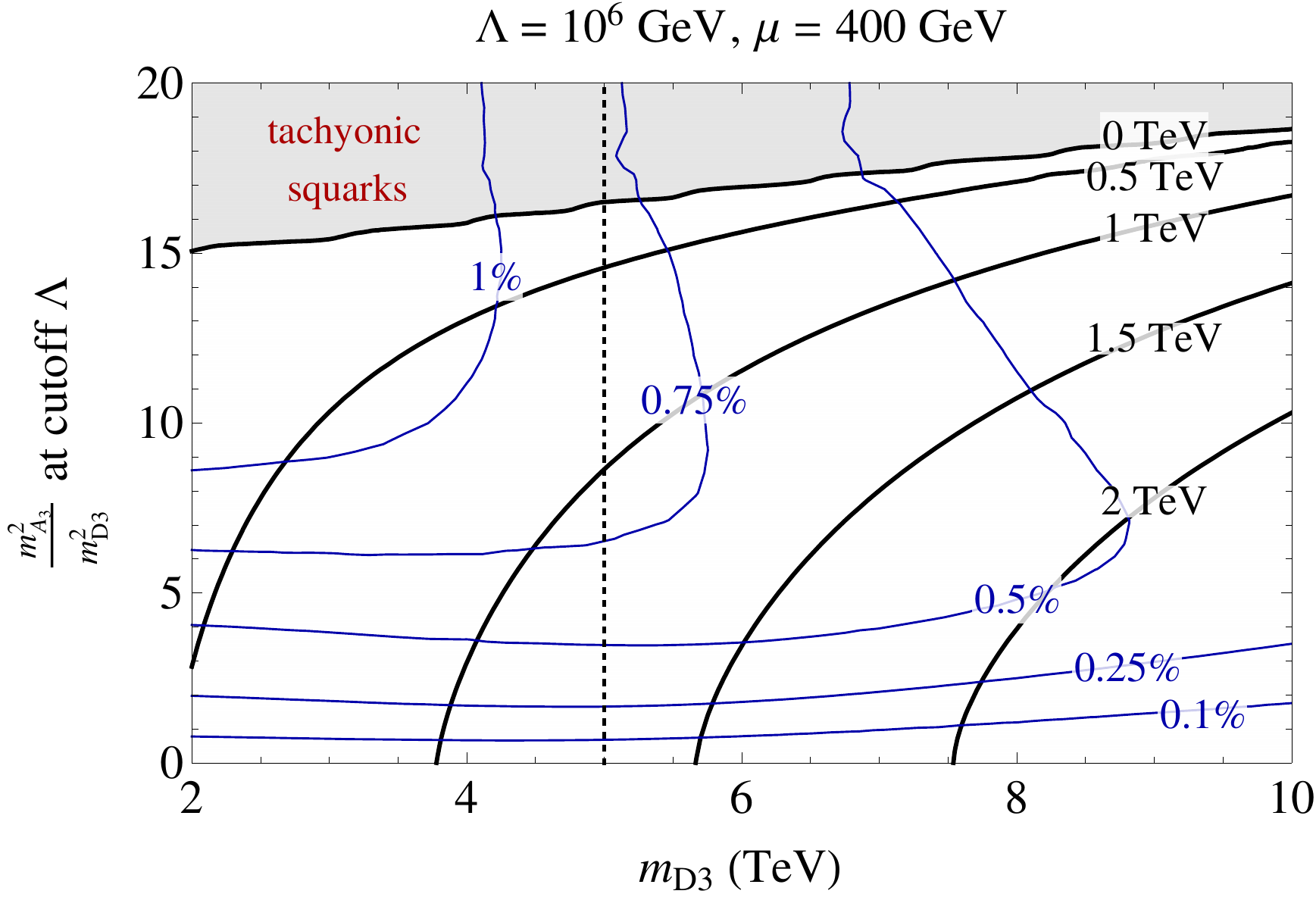}}
\end{minipage}
\vspace{-4mm}
\caption{(a) Squark mass contours (black) as a function of the mediation scale $\Lambda$ and the ratio $\left.m_{A_3}^2\right/m_{D3}^2$ at the scale $\Lambda$, with the physical Dirac gluino mass fixed at 5 TeV and adjoint scalar masses as in eq.~(\ref{eq:masseigenstates}) with $B_{A_i} \approx 0$.  Contours of right-handed slepton masses are shown in purple under assumptions given in the text. The region $\left.m_{A_3}^2\right/m_{D3}^2 > \frac{2\pi}{\alpha_3}-2$ as in the pure $D$-term model of \cite{Benakli:2008pg} is shown in green; note that its contour changes with $\Lambda$ due to eq.~(\ref{eq:alpha3running}). (b) Squark mass (black) and fine tuning (blue) contours as a function of the physical Dirac gluino mass $m_{D3}$ and the ratio $\left.m_{A_3}^2\right/m_{D3}^2$ at the mediation scale $\Lambda = 10^6\gev$. The black dotted line on each figure represent 1D slices of the other figure.}\label{fig:dg}
\end{figure}

These models of Dirac gauginos are not supersoft at higher orders. In particular, if $m_{A_i}^2 \neq 0$, there is a large two-loop contribution to the squark masses of the form \cite{Goodsell:2012fm,Martin:1993zk}\footnote{The first version of \cite{Goodsell:2012fm} differed from eq.~(\ref{eq:twoloopcont}) by a factor of 8/3.}:
\begin{align}
\partial_t m_{\tilde{q}}^2  \simeq \frac{32 \alpha_3^2}{(4\pi)^2} ~ m_{A_3}^2 + \dots .
\label{eq:twoloopcont}
\end{align}
This RG effect drives the squarks tachyonic and can dominate the positive, finite contribution in eq.~(\ref{eq:sfermionmasses}).  This means that the ratio $\left.m_{A_3}^2\right/m_{D3}^2$ at the SUSY-breaking scale $\Lambda$ needs to be sufficiently small
to avoid tachyonic or unacceptably light squarks.\footnote{Electroweak symmetry breaking may also place a bound on the ratio $\left.m_{A_3}^2\right/m_{D3}^2$ in explicit models, because above the approximate threshold scale $\sqrt{m_{D3} m_{A_3}}$, the stop mass is negative when $m_{A_3}^2>0$, giving \textit{positive} contributions to $m_{H_u}^2$.} 
The generic gauge mediated prediction of this ratio---an inverse loop factor---results in tachyons for low and intermediate mediation scales; we show squark mass contours as a function of $\Lambda$ and $\left.m_{A_3}^2\right/m_{D3}^2$ in figure~\ref{fig:dg}.  We also show mass contours of the right-handed sleptons, under the assumption of degenerate adjoint mass terms and $m_{Di}/m_{Dj} = g_i/g_j$ at the scale $\Lambda$.

It may be possible to construct a low-scale model in which $m_{Di}^2 \sim m_{A_i}^2 \sim B_{A_i}$ with the correct relative signs; however, we expect such an arrangement to be accompanied with additional fine tuning given the natural loop suppression of $m_{Di}^2$ vs.~$m_{A_i}^2$.  We quantify the total fine tuning in such a low-scale model in figure~\ref{fig:dg}(b), where we assumed: 
\begin{align}
\text{FT}_\text{total} = \frac{\left.m_{A_3}^2\right/m_{D3}^2}{\pi/\alpha_3}\times \text{FT}_{m_h^2}\left[a_i\right],
\end{align}
with $a_i = \lbrace\mu^2,m_{H_u}^2,m_{D3}^2,m_{A_3}^2\rbrace$.  We take $\mu~=~400~\gev$ in figure~\ref{fig:dg}(b) to satisfy the bounds on higgsinos in models of low-scale gauge mediation. We find that low-scale models of Dirac gauginos are tuned to at least 1\%.\footnote{If one considers only the observable $m_h^2$ and not $\left.m_{A_3}^2\right/m_{D3}^2$, the tuning is at best 3.5\%, e.g.~for the parameters $m_{D3}=4\tev$, $\left.m_{A_3}^2\right/m_{D3}^2 \approx 0$, $\mu = 400\gev$, and $\Lambda = 10^6\gev$.}

For sufficiently high mediation scales, $\Lambda \gtrsim 10^{15} \gev$, the squark tachyon problem can be avoided more easily.  This is because $m_{A_3}^2$ is irrelevant in the IR compared to $m_{D3}^2$ and $B_{A_3}$; at one-loop order with preserved $R$-symmetry, the $\beta$-functions are:
\begin{align}
\partial_t m_{A_3}^2 = 0,\qquad \frac{\partial_t m_{D3}^2}{m_{D3}^2} = \frac{\partial_t B_{A_3}}{B_{A_3}} =  -\frac{3\alpha}{\pi}.\label{eq:massrunning1}
\end{align}
These relations imply that the theory flows to the supersoft limit
($|m_{D3}^2|,|B_{A_3}|\gg| m_{A_3}^2|$) in the IR, with only moderate
contributions from $m_{A_3}^2$to the squark masses
(eq.~(\ref{eq:twoloopcont})).  To avoid tachyonic octet scalars at low
scales, it is sufficient that $4m_{D3}^2 > -B_{A_3} > 0$ at the
mediation scale (eqs.~(\ref{eq:masseigenstates}) and
(\ref{eq:massrunning1})).  However, RG flow from
e.g.~$\Lambda\sim10^{16} \gev$ with universal gaugino mass boundary
conditions creates a factor of $\gtrsim 15$ splitting between the
gluino and bino masses, yielding $\gtrsim 5\tev$ squark masses if
right-handed sleptons are above $\sim 100\gev$.  A combination of $F$-
and $D$-terms can give $\mathcal{O}(100\gev)$ masses to the sleptons
and $\mathcal{O}(1\tev)$ masses to the squarks at low scales.
However, a typical feature of high-scale mediation models is that the
octet superfield $A_3$ acquires an $R$-symmetry-breaking
superpotential mass $M_{A_3}$ via gravity mediation.
 This mass term is by far the most relevant operator in the IR: $\left.\partial_t
  M_{A_3}\right/M_{A_3} \simeq -6\alpha_3/\pi$.  When dominant, it
yields a light gluino mass eigenstate of mass
$\sim\left.m^2_{D3}\right/M_{A_3}$ at low scales through a seesaw
mechanism.  Hence the natural separation between the stop and Dirac
gluino mass is destroyed, so naturalness is again spoiled due to the
strong LHC bounds on the gluino.

We did not touch upon other model-building challenges -- such as avoiding a vev for the $SU(2)$ adjoint scalar and generating large enough effective $\mu$, $B\mu$, and Higgs quartic coupling --  which have been covered extensively in bottom-up studies \cite{Benakli:2012cy,Benakli:2011kz}.  In particular, one additional contribution to the tuning of the Higgs mass could come from operators such as $\mathcal{W} \supset H_u A_{1,2} H_d$ especially if $A_1$ and $A_2$ are heavy \cite{Goodsell:2012fm}.  These obstacles notwithstanding, it is clear from figure~\ref{fig:dg} and the above discussion that the colored sector---controlled by the Dirac gluino and colored octet mass parameters---must be very peculiar to obtain a viable, let alone natural, model of Dirac gauginos.


\section{Summary}
\label{sec:conclusion}

The previous sections show that naturalness of the EW scale is being
severely challenged by the LHC. Even if the Higgs sector is extended
to explain the measured value of 125 GeV, as in the NMSSM, the
tuning imposed by direct searches of colored sparticles is at
the percent level, not much improved compared to the ordinary MSSM with large A-terms (see table~\ref{tab:sumtun}). To have a chance of improving the situation we need to go beyond the minimal model and look for mechanisms that explain the absence of signals at the LHC. Unfortunately, all the alternatives proposed so far in the literature seem to fail mainly because of the strong bounds on the gluino mass. 

Because of its large color charge, the gluino is at the same time the most copiously produced SUSY particle at the LHC and a strong attractor in the RGEs for the stop and consequently for the Higgs. In most models, the gluino bounds are at or above the TeV level, which translates to a few percent tuning for split-family or even baryonic RPV models. In the latter case, the bounds on the gluino remain strong in part because the missing energy removed by RPV decay of the LSP is substituted by the leptons and missing energy that are produced from top production from the cascade decay: a natural spectrum should always contain light stops and sbottoms, making top production hard to avoid.

\begin{table}[t]
\begin{center}
\begin{tabular}{| c || c |}
\hline
Model & FT \\ \hline \hline
Mini-Split & $\lesssim$0.05\% \\
MSSM & 0.3--1\% \\
NMSSM & 2\% \\ 
Split Families & 3\% \\ 
bRPV & 2--3\% \\ 
Dirac Gauginos & ($\lesssim$1\%). \\  \hline
\end{tabular}
\end{center}
\caption{\label{tab:sumtun} Summary of the level of fine-tuning in the representative models studied in this paper compared to the `unnatural' Mini-Split models \cite{Arvanitaki:2012ps,ArkaniHamed:2012gw}.}
\end{table}

One may hope to relax the tuning in models of Dirac gauginos where the
gluino can be out of reach by the LHC and does not contribute with log
enhancement to the stop and Higgs masses because of supersoftness.
Dirac gaugino models, however, have several challenges. First, the unification of gauge couplings is no longer automatic---the only experimental success of SUSY is lost. Second, the supersoft property is not ensured by a symmetry. On the contrary, in all known calculable models, supersoftness is badly broken and has to be recovered by cancellation, which makes the tuning below the 1\% level. Successful model building with  Dirac gauginos also has to deal with large RG corrections which may produce tachyons or big hierarchies in the spectrum that bring additional tension between naturalness and LHC bounds.  Unless all these issues are addressed in a natural manner, one cannot claim that Dirac gauginos are a solution to tuning in SUSY.

In conclusion, while going beyond the MSSM may ameliorate naturalness, the improvement seems mild and only bring the tuning from 0.5-1\% to the few percent level. Considering the required efforts in model building and the
increased complexity of the resulting models it is natural to wonder whether such improvements are real. 
Whenever naturalness is ameliorated with an increase in complexity, there are hidden ``unspeakable'' tunings that are not captured by the standard formula but should be taken into account---every ingredient added and every parameter chosen to hide SUSY from us should in principle be paid in the tuning bill. Furthermore, since these types of tunings exist only to explain the absence of signals, they cannot even be explained with anthropic arguments and for this reason are qualitatively different, and worse, than the usual tuning of the EW scale.

There are several assumptions that come into the work presented above and warrant further discussion:

\begin{itemize}
\item We use the same measure of tuning and assume gauge mediated SUSY breaking (GMSB) for all the models discussed in order to compute the tuning reliably and guarantee gauge coupling unification as well as a natural solution to the SUSY-flavor problem. The common setup also allows for a fair and robust comparison of the level of the tuning among the different models. 

\item In addition to the gauge mediated contributions at the messenger
  scale, we allow for extra contributions to the Higgs-stop sector as
  suggested by models that address the $\mu$-$B_\mu$ problem. Assuming
  GMSB allows us to lower the mediation scale of the theory, reducing
  effects from RG running. On the other hand, the mediation scale
  cannot be pushed all the way down to the TeV scale without losing
  perturbativity, thus still leaving large RGE corrections to $m_{H_u}$ and $m_{\tilde t}$. \\
  These limitations may be avoided in deconstructed models of gaugino
  mediation where a no-scale spectrum can be generated at low scales
  while still compatible with unification
  \cite{Cheng:2001an,Csaki:2001em}. Since the scalar boundary
  conditions vanish and the RGE effect is minimized by the low
  effective scale (which can even be below $10\tev$), the tuning in
  these models may be relaxed above 10\%. No-scale boundary conditions
  may however be troublesome because they lead to light right-handed
  sleptons---imposing the experimental bounds on these generically
  leads to stops above the TeV scale~\cite{DeSimone:2008gm},
  reintroducing tuning.  Also perturbative unification and the
  $\mu$-$B_\mu$ problem require extra model building in this class of
  models.

\item Except for the MSSM example, we neglect any tuning required to fix the Higgs quartic coupling to its experimental value, and instead focus only on the tuning coming from the direct experimental bounds on the superpartners. Most known models to raise the Higgs mass are not completely natural and produce extra contributions to the tuning. Similarly, we neglect possible tuning hidden in the UV model addressing the $\mu$-$B_\mu$ problem and in the SUSY breaking sector generating the right gauge mediated setup. From this point of view the overall values of the tunings in table~\ref{tab:sumtun} should be interpreted as an optimistic upper bound to the actual amount of tuning.
\item We did not consider combining models (e.g.~split families or Dirac gauginos in bRPV). Even though such tricks may improve the canonical tuning, the increase in complexity (for the mere purpose of hiding SUSY from experimental searches) corresponds to an increase of tuning in theory space. While it is reasonable to expect that a property of the theory makes SUSY harder to be found, it is difficult to believe that this is achieved by combinations of effects acting coherently. Such a conspiracy would not be different from the canonical tuning.
\item It is clear that several of the experimental bounds we present are model dependent or rely on simplified models. In the case of simplified models, unless the branching ratios through the specific channels are significantly modified, the bounds may be only slightly relaxed and do not make a difference in the tuning. In the case of bRPV, where we have considered specific spectra when applying the experimental analysis, we again do not expect a significant change when other models are considered. Since natural SUSY spectra should include light stops and higgsinos, we generically expect top production which guarantees leptons and missing energy from the $W$s and $Z$s produced in cascade decays.
\end{itemize}

Hiding SUSY at the LHC is very different from, for example, suppressing SUSY contributions to flavor observables. In order to suppress flavor violation, it is relatively easy to arrange a gauge-mediated model to give universal squark masses---any non-universality generated by RG flow is proportional to the yukawa couplings themselves so it is automatically minimally flavor violating. In contrast, there is no corresponding symmetry for ``hiding" at the LHC: placing superpartners just at the right masses with just the right mass splittings is, by definition, not an RG-invariant statement and these spectra require very special boundary conditions in the UV. The associated tuning is exacerbated by the presence of IR fixed relations that tend to pull the spectrum away from the one required to hide SUSY. Such tuning is qualitatively worse than the standard tuning of the EW scale as it cannot even be explained anthropically.
Models of this kind include compressed SUSY spectra where all jets or leptons are very soft or bRPV spectra which are squeezed to reduce top production.

We thus find that LHC searches for colored sparticles are pushing SUSY in the regime of percent level tuning, despite model-building efforts to explain the absence of sparticles at the LHC. 
The existence of tuning in the EW scale is by now a fact. What is
still under debate is whether such a tuning may be interpreted as an
accidental cancellation in the fundamental theory, which may still
explain the smallness of the EW scale dynamically, or if it is the
first signal that the smallness of the EW scale is the result of
environmental selection, which does not require the presence of new
light states. While the level of tuning required to distinguish the
two possibilities is quite subjective, the numbers in
table~\ref{tab:sumtun} should start making us quite uncomfortable with the first option.

Apart from the cosmological constant problem, this is the first time
that effective field theory estimates fail so dramatically.  Examples
from QCD and nuclear physics (such as the size of the scattering
length of certain di-nucleon systems) are sometimes used as arguments
for the plausibility of percent-level tunings in nature. However, we
do not think that such arguments can be used in support of
naturalness for the EW scale.  First, the examples referred above are
often affected by QCD uncertainties, which makes a fair estimate of
the tuning hard. Second, even accepting the most pessimistic estimates
(which are at the few percent level), tunings appear only after having
looked at many QCD observables and its relevance is washed out after
taking into account the look-elsewhere effect. Nobody would be
surprised if, after having discovered SUSY and measured its spectrum,
one of the observables would appear to be tuned at the few percent
level. On the contrary, finding a gap above the EW-scale appear as
surprising as it would have been to find a gap between the pion and
the other QCD resonances which is bigger than the natural one derived from the
$\pi^+$-$\pi^0$ mass difference.

Percent cancellations are beyond the threshold of our tolerance and
make us wonder: will a discovery of SUSY particles in the second run of
the LHC be a true triumph of naturalness or a confirmation of its
failure?

\acknowledgments{}

\noindent We warmly thank Savas Dimopoulos and Tony Gherghetta for
collaboration in the early stage of this paper and for their valuable
feedback throughout the completion of the work.  We are also grateful
to Tim Cohen, Nathaniel Craig, Mark Goodsell, Sonia El Hedri, Kiel Howe, David
E. Kaplan, John March-Russell, Riccardo Rattazzi, Prashant Saraswat,
Yuri Shirman, and Florian Staub for many useful comments and
discussions. 

We thank the CERN Theory Group for their hospitality during the
completion of this work, and we also thank the Galileo Galilei
Institute for Theoretical Physics for the hospitality and the INFN for
partial support during the completion of this work.  We acknowledge
the help from FarmShare, Stanford's shared computing environment,
which made MC event generation and analysis possible. 

This work was supported in part by ERC grant BSMOXFORD no. 228169.  MB
is supported in part by the NSF Graduate Research Fellowship under
Grant No. DGE-1147470. XH is supported in part by the Stanford
Humanities and Sciences Graduate Fellowship.


\appendix
\section{Taking into account correlations between different sources of tuning}\label{app:tuning}
For the computation of the tuning in the different models we use the
definitions in eqs.~(\ref{eq:deftune1}) and
(\ref{eq:deftune2}). Unlike in most of the literature where the tuning
is computed taking only into account the dominant contribution in the
sum in eq.~(\ref{eq:deftune1}), we include all the contributions in
quadrature in order to reproduce the expected growth of the tuning
with the square root of the number of parameters when they equally
contribute to a given observable.

For the reasons explained in section~\ref{sec:QN}, for the tuning of the EW vev we always use the Higgs formula
\begin{equation}
m_h^2=-2(|\mu|^2+m_{H_u}^2)\,,
\end{equation}
which gives a fair estimate of the tuning in all cases.
The dependence on the fundamental parameters enters through $m_{H_u}^2$. In particular we consider the dependence with respect to: the $\mu$ term, the gluino mass at the messenger scale, the corrections to the gauge mediated boundary values of $m_{H_u}^2$ and $m_{\tilde t}^2$ at the messenger scale and other high-scale parameters the models may depend on, such as $A$-terms in the MSSM model or the extra messenger sector in the split-family model.

Unless specified differently, the total tuning includes the tuning against the EW vev and the stop mass.
Since these two quantities depend on the same set of UV parameters
they are not completely independent and we have to make sure not to
overestimate the tuning.\footnote{We thank David Pinner and Josh Ruderman for raising this potential issue.} For this purpose we employ the following formula for the total tuning:
\begin{equation} \label{eq:corrtune}
\text{FT}={\rm Min} \left[{\rm Min}\left(1,\text{FT}_{m_h^2} \right){\rm Min}\left(1,\frac{\text{FT}_{m_{\tilde t}^2}}{\sin\theta} \right) ,
{\rm Min}\left(1,\text{FT}_{m_{\tilde t}^2} \right){\rm Min}\left(1,\frac{\text{FT}_{m_{h}^2}}{\sin\theta} \right) \right] \,,
\end{equation}
where
\begin{align}
\sin\theta &\equiv \frac{\left|\vec{v}_{m_h^2} \times \vec{v}_{m_{\tilde t}^2}\right|}{\left|\vec{v}_{m_h^2}\right| \left|\vec{v}_{m_{\tilde t}^2}\right|}\,, \quad
{v}^i_x \equiv \frac{\partial \log x}{\partial \log a_i}\,, \quad
\text{FT}_x \equiv \frac{1}{|\vec{v}_{x}|} \,.
\end{align}
This formula interpolates between 
\begin{equation}
\text{FT}=\text{FT}_{m_h^2} \cdot \text{FT}_{m_{\tilde t}^2}\,,
\end{equation}
when the two tuning are completely independent ($\sin\theta=1$) and 
\begin{equation}
\text{FT}={\rm Min}\left[ \text{FT}_{m_h^2}, \text{FT}_{m_{\tilde t}^2} \right] \,,
\end{equation}
when they are completely dependent ($\sin\theta=0$).

Having a large correlation implies that by tuning the stop to be light against the gluino, the Higgs mass is automatically relieved by the largest contribution and does not require additional tuning. For high-scale messengers this is not the case because for most of the running the stop mass is large due to the gluino attraction and gives a significant contribution to the $m_{H_u}^2$. The correlation is also not important in low-scale gauge mediated models because of the extra contributions to $m_{H_u}^2$ and $m_{\tilde t}^2$ at the messenger scale. In fact in all our examples we checked that the effect from the correlation is never significant and treating the two tunings as independent gives a good estimate of the overall tuning.

\section{Split Families details}
\label{sec:u1p-appendix}

Here we explain the split family model in more detail.  All particle
content receives soft SUSY breaking masses via gauge mediation from
vector-like $\boldsymbol{5+\bar{5}}$ messenger pairs $(D' +
\bar{D}',L' + \bar{L}')$. The first two generations receive additional
soft SUSY breaking masses through a pair of messengers $N+\bar{N}$
which are charged only under the new $U(1)'$. The messengers
communicate the effects of SUSY breaking parameterized by spurions
$X_D = m_D + F \theta^{2}$ and $X_N = m_N + F \theta^2$:
\begin{equation}
  \mathcal{W}_{\text{mess}} = X_D(D'\bar{D}' + L'\bar{L}') + X_N N \bar{N}.
\end{equation}

Since the third generation is not charged under $U(1)'$, the flavor
structure of our model is non-trivial. The Yukawa terms cannot contain
mixing between the first two and third generations:
\begin{equation}
  \begin{split}
    \label{eq:mssm_superpotential}    
    \mathcal{W}_{\text{MSSM}} =& y_{ij}^{u}H_uQ_iU_j + y^u_3H_uQ_3U_3 +    y^d_{ij}H_dQ_iD_j    + y^d_3 H_dQ_3D_3 +\\
    &y^e_{ij}H_dL_iE_j+y^e_3H_dL_3E_3 + \mu H_uH_d \quad (i,j = 1,2) \,.
  \end{split}
\end{equation}
The spontaneous breaking of $U(1)'$ generates the mixing
between $D_3$ and messenger $\bar{D}''$ from superpotential
\begin{equation}
  \label{eq:u1p-superpotential-mixing}
  \mathcal{W} \supset \lambda_{i}H_dQ_iD'' + \Phi \bar{D}'' D_3 + m_{D''}D''
  \bar{D}'' + V(\Phi)_{U(1)'\text{-breaking}} \quad\quad (i = 1,2).  
\end{equation}
Integrating out the messengers $D''$
and $\bar{D}''$ generates the required Yukawa mixing:
\begin{equation}
  \label{eq:mixingintegrated}
  \mathcal{W}_{\text{mixing}} =  \lambda_{i}\frac{\vev{\Phi}}{m_{D''}} H_d Q_i D_3,
\end{equation}
with $\lambda_{i}\sim \mathcal{O}(1)$. Note that messengers
$D'',\bar{D}''$ cannot couple to the SUSY breaking spurions; otherwise
unacceptably large $A$-terms and off-diagonal soft masses will be
generated, resulting in large flavor violation. Forbidding this
coupling can be accomplished by assigning appropriate $U(1)'$ charges
to $D'$, $D''$, and the SUSY breaking spurions.

To summarize, there are $4$ scales: $m_{D''} > m_{D} > m_N \gtrsim
\vev{\Phi}$. The ratio $\vev{\Phi}/m_{D''}$ sets the magnitude of the Yukawa
mixing between the third and first/second generations; $m_D/m_N$ sets
the mass splitting between the sfermions. $m_{N}$ sets the scale of
mediation to the first two generations. We take $m_N \gtrsim \vev{\Phi}$ so
that the gauge mediation from $U(1)'$ gauge bosons is not
suppressed. 

\subsection*{Contribution to flavor observables}
\label{sec:contr-flav-observ}

The superpotential given in eqs. (\ref{eq:mssm_superpotential}) and
(\ref{eq:mixingintegrated}) is not in field basis convenient for comparison
with the Standard Model Yukawa terms; in particular, there is no
$Q_3D_1$ or $Q_3D_2$ mixing. Therefore to study the flavor structure,
we need to match our parameters $(y^u_{ij}, y^d_{ij},y^u_3,y^d_3,
\lambda_{i})$ to the Yukawas and CKM elements and rotate to the new
basis accordingly. Only 10 (real) parameters are
physical,\footnote{$(y^u_{ij}, y^d_{ij},y^u_3,y^d_3, \lambda_{i})$
  contains 12 complex, or 24 real parameters. Unitary rotations of the
  first two generations render $4\times 3$ of them unphysical, and 3
  additional phases can be absorbed into the third generation
  fields. Adding back the overall conserved baryon number, there are
  $24-12-3+1=10$ real physical parameters left. } corresponding exactly
to the 10 degrees of freedom (6 quark masses, 3 CKM angles, 1 CP
phase) in the Standard Model. The field rotation disturbs the
flavor-diagonal nature of gauge mediation and therefore can induce
potentially dangerous flavor violation (see e.g.~\cite{Mescia:2012fg,Giudice:2008uk,Barbieri:2010pd}).

In our toy model, however, such effects are small; the smallness can
be understood best in a basis where the up sector is diagonal. In this
basis, the three $D_i$ need to be rotated from a basis where the down
sector mass matrix is 0 in the $(3,1)$ and $(3,2)$ entries, to a basis
where the mass matrix is $V_{\text{CKM}} D$, where $D$ is a diagonal
matrix with $y_d$, $y_s$, and $y_b$ on the diagonal. Numerically, the
$(3,1)$ and $(3,2)$ entries of $V_{\text{CKM}} D$ are much smaller
than the other entries, therefore requiring only a small rotation away
from the gauge mediation basis. We verified with \texttt{SUSY\_FLAVOR
  v2.02} \cite{Crivellin:2012jv} that the additional contribution to
sensitive flavor observables agree with the SM prediction within
theoretical error. We checked that already for first two generation
squarks at 3~TeV and above the corrections to $\varepsilon_K$ and
other flavor observables are negligible.

\section{RPV details}\label{app:rpv}

Here we discuss in more detail the `toy model' presented in
Section~\ref{rpv_model}; in the following subsections we provide
further details of limits on UDD couplings and discuss our procedure
for event generation and recasting of experimental analyses to set
limits on the model.

In order to preserve the success of unification in the MSSM while
introducing baryon but not lepton number violation, we use a
solution already in place for the Higgs doublet-triplet splitting
problem: orbifold GUTs \cite{Hall:2001pg}. To generate the RPV
couplings, consider an orbifold with $SU(5)$ in the bulk, and
$SU(3)\times SU(2)\times U(1)$ on the IR brane ($z=0$). The matter
content has varying profiles in the bulk.  The Higgses are on the
IR brane, leading to a different yukawa hierarchy in the up and down sectors,
\begin{equation}
y^u_i= (\left.Q_i\right|_{z=0})^2 = (\left.U_i\right|_{z=0}
)^2\quad \mathrm{while}\quad y^d_i=(\left. Q_i\right|_{z=0}) (\left. D_i\right|_{z=0}).\label{eq:2}
\end{equation}

\begin{figure}[h]
\begin{center}
\hspace{3cm}   \includegraphics[trim = 0mm 0mm 0mm 0mm, clip, width
      = 0.6\textwidth]{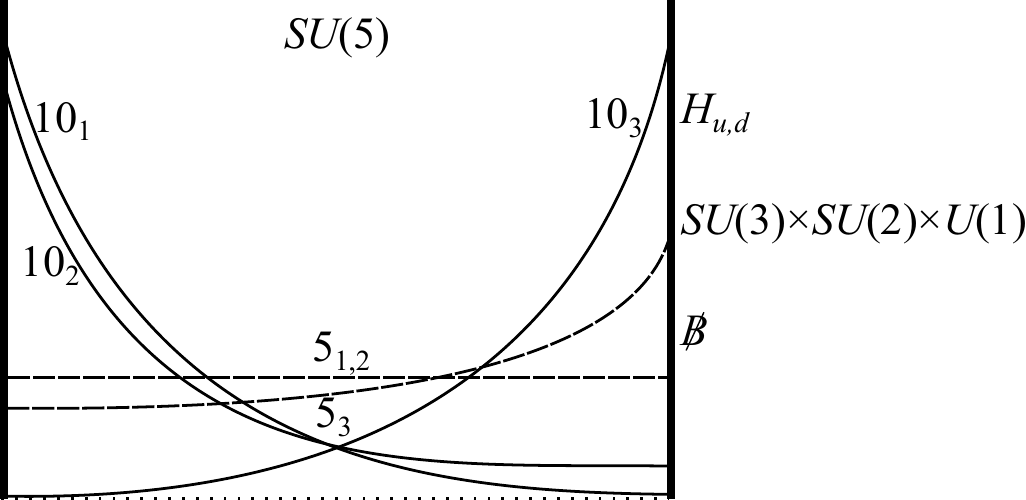}
\vspace{-2mm}
\caption{Toy orbifold model for bRPV.} \label{fig:rpvmodel} \vspace{-2mm}
   \end{center}
\end{figure}

On the IR brane we introduce heavy $D'$ and $\Phi$ fields charged under baryon number
and with $B$-invariant superpotential couplings
\begin{equation}
\left.\mathcal{W}_{RPV}\right|_{z=0} \supset \kappa_i U_i D' D' +
\overline\kappa_j \Phi D_j \bar D' + M_{D}D' \bar D' \,,
\end{equation}
the baryon number is then broken spontaneously after the field $\Phi$ gets a vev.{\footnote {There may be a light axion that is associated with the breaking of baryon number. Its exact mass depends on the details of the UV model and its axion decay constant is close to the GUT scale, so it doesn't introduce significant experimental constraints.}}
Since B-breaking is
introduced on the IR brane, leptonic partners of down-type fields are
not required to preserve unification, and do not induce lepton number
violation. Integrating out the $D'$ field then gives a baryon number
violating term,
\begin{equation}
\mathcal{W}_{\slashed{B}} =\frac
{\vev{\Phi}^2}{M_{D}^2}\left. \kappa_i \overline\kappa_j
  \overline\kappa_k \, U_iD_jD_k\right|_{z=0}  \,,
\end{equation}
we take $\kappa_i \sim \overline\kappa_j\sim\mathcal{O}(1)$, resulting
in a predictive pattern of RPV couplings,
\begin{equation}
\label{eq:rpv}
\lambda''_{ijk}\simeq \frac{\vev{\Phi}^2}{M_{D}^2}\left.(U_i\,D_j\,D_k)\right|_{z=0}
= \frac{\vev{\Phi}^2}{M_{D}^2}\sqrt{y^u_i} \, \frac{y^d_j}{\sqrt{y^u_j}}
 \frac{y^d_k}{\sqrt{y^u_k}}.
\end{equation}
SUSY breaking is mediated via doublet-triplet split gauge mediation
such that the ratio of the wino to the gluino mass may be allowed to
vary;
we find, however, that a heavy wino does not significantly reduce
experimental bounds at the expense of increased tuning from an
additional parameter.

\subsection*{RPV coupling constraints}
One of the most constraining limits on $\lambda''_{ijk}$ comes from
$n$-$\overline{n}$ oscillations: the tree level process puts limits on
$\lambda''_{11k}$. Because of the left-right insertions in this
process,  a more constraining limit on $\lambda''_{112}$ results from
decays $^{16}O~\rightarrow~^{14}C K^+ K^+$, through $n-\Xi$
oscillations. For a decay lifetime $\tau > 10^{31}$s, the limit is
\cite{Barbieri:1985ty,Goity:1994dq},
\begin{equation}
\lambda''_{112} \lesssim 10^{-5} \left(\frac{m_{\tilde
g}}{600\gev}\right)^{1/2} \left(\frac{m_{\tilde
d_R}}{600\gev}\right)^2 
\left(\frac{10^{-6}\gev^6}{\bra{n} u_rd_ru_rd_rs_rs_r\ket{\Xi}}\right)^{1/2}  \,.
\end{equation}
These limits are subject to large nuclear uncertainties; in addition,
$n$-$\overline{n}$ oscillations place a limit of the same order of
magnitude $\lambda''_{121}$ where one of the quarks participating in
the process is a strange quark which makes up about $10\%$ of the
neutron at zero momentum.

Loop-level $n$-$\overline{n}$ oscillations also put a constraint on
third generation RPV couplings $\lambda''_{3jk}$,
\begin{align}
\lambda''_{321} &<  0.15 \left(\frac{M_2}{600\gev}\right)^{1/2} \left(\frac{m_{\tilde
q_L}}{200\gev}\right)^2 \left(\frac{m_{\tilde
q_R}}{200\gev}\right)^2\left(\frac{\left(600\gev \right)^2}{A_{LR}^t
A_{LR}^s}\right) \left(\frac{10^{-4}\gev^6}{|\psi(0)|^4}\right)^{1/2}\,, \\
\lambda''_{331} &<  0.3 \left(\frac{M_2}{600\gev}\right)^{1/2} \left(\frac{m_{\tilde
q_L}}{600\gev}\right)^2 \left(\frac{m_{\tilde q_R}}{600\gev}\right)^2\left(\frac{\left(600\gev \right)^2}{A_{LR}^t A_{LR}^b}\right) \left(\frac{10^{-4}\gev^6}{|\psi(0)|^4}\right)^{1/2} \,.
\end{align}
There are also limits on products of $\lambda''$
\cite{Barbier:2004ez}; the most constraining limits result from rare
$B$ decays and $K$-$\bar K$ oscillations, and provide limits on
second-generation couplings, but are less constraining for first and
third generations.

\begin{equation}
  \label{eq:1}
|\lambda_{ijk}^{''*}   \lambda_{ijk}''|^{1/2} \lesssim 10^{-1} \mathrm{-} 10^{-2}\,.
\end{equation}

Taking into account all of the above constraints, the strongest limits
on the model which has a square root of yukawa coupling hierarchy come
from the first generation, $\lambda''_{1jk}\lesssim 10^{-5}$ for squarks at
$600\gev$. This is only an order of magnitude bound due to the large
nuclear uncertainties in this limit; we pick the maximum allowed RPV
couplings to study the phenomenology of the model \eqref{eq:5}.

\subsection*{Simulation details}
\begin{figure}[t]
\begin{center}
    \subfigure[]{\includegraphics[trim = 0mm 0mm 0mm 0mm, clip, width
      = 0.49\textwidth]{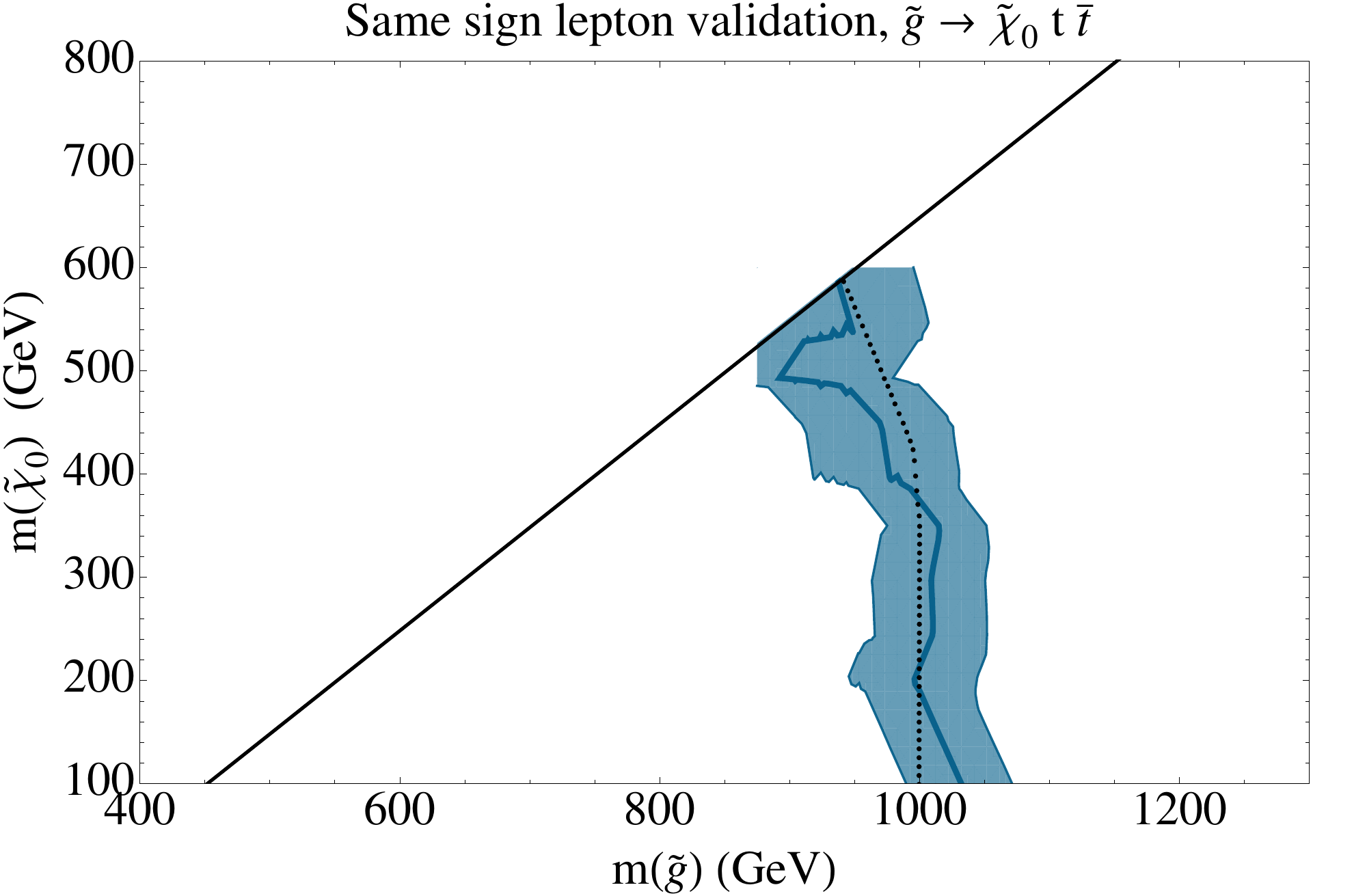}\label{fig:validationL}}
\hspace{-2mm}
    \subfigure[]{\includegraphics[trim = 0mm 0mm 0mm 0mm, clip, width
      = 0.49\textwidth]{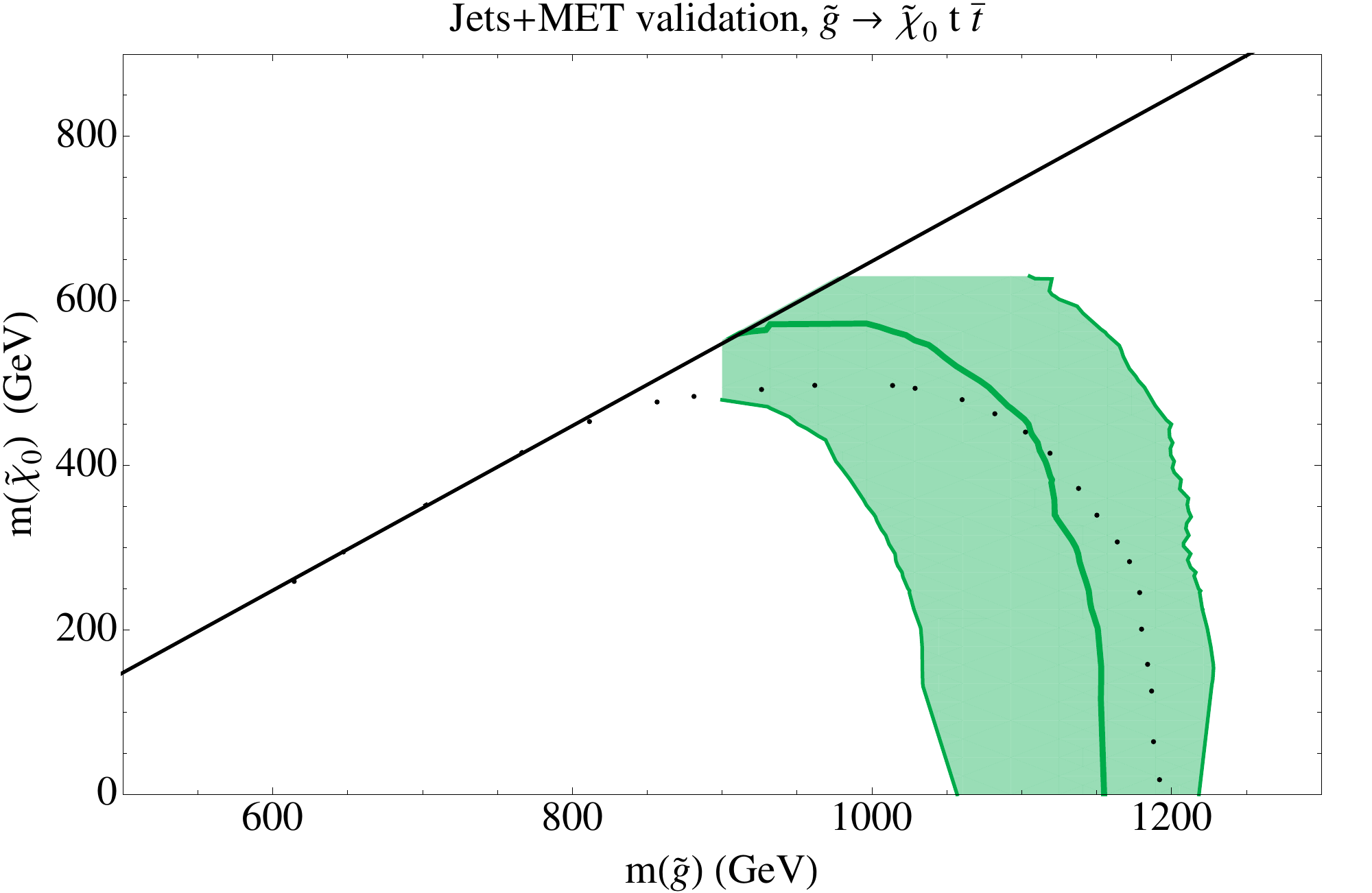}\label{fig:validationJ}}
\vspace{-2mm}
\caption{ Validation of search implementations for (a) CMS SS
  dileptons and b-jets \cite{Chatrchyan:2012paa} and (b) ATLAS 7-10
  jets and MET \cite{ATLAS-CONF-2013-054}. The experimental limit is
  indicated with the dotted line, and our limit has been rescaled to
  match the observed bound by (a) $2.4 \pm 0.6$ and (b) $1.1^{+ 1.6}_{
    - 0.5}$. The leptonic search matches quite well after the
  rescaling, which may be a result of overestimating lepton
  efficiencies; the jet search is an overestimate in the squeezed
  region, and we take this into account by introducing a large
  uncertainty. } \label{fig:validation} \vspace{-2mm}
   \end{center}
\end{figure}

Given that the current LHC limits as applied to RPV models are
limited, we recast the most relevant searches to set limits on the
model described in Section~\ref{rpv_model}. We summarize our procedure
below. Our bounds tend to be stronger than other re-castings of the
limits which have been done in the literature; this is due either to
older searches with smaller data sets
\cite{Asano:2012gj,Allanach:2012vj}, or assumptions of restricted
production channels \cite{Durieux:2013uqa,Berger:2013sir}. We find,
for example, that our same-sign lepton search recasting sets a limit
of $M_3 >850\gev$ for the gluino pair-production channel,
which matches the limit derived by other authors
\cite{Berger:2013sir}.

We create a grid in the gluino-stop plane in order to study the
constraints. We simulate the hard scattering LHC processes at
$\sqrt{s} = 8\tev$ for the leading squark-squark, squark-gluino, and
gluino-gluino production channels \texttt{MadGraph5 v1.5.7}
\cite{Alwall2011}. For each parameter point we compute the low-energy
spectrum using \texttt{SOFTSUSY v3.3.4} \cite{Allanach:2001kg}; we
calculate the branching ratios, including the RPV couplings according
to the pattern in eq.~\eqref{eq:5} and assuming a Higgs mass of
$m_h=125\gev$ using \texttt{BRIDGE v2.24} \cite{Meade:2007js} and
\texttt{MadGraph5}. We confirm the branching ratios with
\texttt{SDECAY v1.3} \cite{Muhlleitner:2003vg} and find agreement
between the two methods within $5\%$. We compute NLO cross-sections
for the processes in \texttt{Prospino v2.1} \cite{Beenakker:1996ed}
and decay and shower the complete spectrum using \texttt{Pythia
  v8.175} \cite{Sjostrand:2007gs}. Finally, we use the detector
simulator \texttt{Delphes v3.0.5} \cite{Ovyn:2009tx} with jet radius,
and lepton and b-tag efficiencies as specified in the corresponding
CMS and ATLAS analyses
\cite{ATLAS-CONF-2013-054,Chatrchyan:2012paa}. 

To check our analysis
against experimental searches, we use our pipeline of
\texttt{MadGraph5}, \texttt{Pythia}, and \texttt{Delphes} on an
example model of $p\,p\rightarrow \tilde{g}\tilde{g},\,\, \tilde g
\rightarrow t\, \overline{t} \tilde\chi^0$ through an off-shell
stop. We find that the limits of our analysis are in good agreement
with those set by the experimental collaborations, and we rescale our
efficiencies by factor of $2.4 \pm 0.6$ and $1.1^{+ 1.6}_{ - 0.5}$ in
the leptonic search and the jets search, respectively, to match the
experimental results (see figure~\ref{fig:validation}), which
introduces an uncertainty in our limits of $150\gev$ in the leptons
and $200\gev$ in the jet search. We use the same central values and
uncertainty bands as in the validation plots to set limits on the RPV
spectrum in figure~\ref{fig:naturalrpv}. It would be very interesting
to see an experimental analysis of these searches in the context of
RPV, as well as a wider range of experimental recastings for
simplified models with RPV decays.

\bibliography{Naturalness}
\bibliographystyle{JHEP}

\end{document}